\documentclass[twocolumn,showpacs,preprintnumbers,amsmath,amssymb,floatfix]{revtex4}

\usepackage{graphicx}
\usepackage{dcolumn}

\newcommand{\ee}{\end{eqnarray}}
\newcommand {\be}[1]{\begin{eqnarray} \mbox{$\label{#1}$}  }

\newcommand{\tpl}{\frac{2\pi}{L}}
\newcommand{\pn}{\frac{\Phi}{\Phi_0}}

\begin{document}

\title{Quantum rings for beginners: \\ 
Energy spectra and persistent currents}

\author{S. Viefers$^1$, P. Koskinen$^2$, P. Singha Deo$^3$, and M. Manninen$^2$}
\address{$^1$Department of Physics, University of Oslo, P.O. Box 1048 
Blindern, N-0316 Oslo, Norway}
\address{$^2$Department of Physics, University of Jyv\"askyl\"a,
FIN-40351 Jyv\"askyl\"a, Finland}
\address{$^3$S.N. Bose National Centre for Basic Sciences, JD Block, 
Sector III, Salt Lake City, Kolkata 98, India}

\date{\today}

\begin{abstract}
Theoretical approaches to one-dimensional and quasi-one-dimensional 
quantum rings with a few electrons are reviewed. 
Discrete Hubbard-type models and continuum models are shown to give similar 
results governed by the special features of the one-dimensionality.
The energy spectrum of the many-body states can be described
by a rotation-vibration spectrum of a 'Wigner molecule' of 
'localized' electrons, combined with the spin-state determined from an
effective antiferromagnetic Heisenberg Hamiltonian.
The persistent current as a function of the magnetic flux through the ring
shows periodic oscillations arising from the 'rigid rotation' of the 
electron ring. For polarized electrons the periodicity 
of the oscillations is always the flux quantum $\Phi_0$.
For nonpolarized electrons the periodicity depends on the 
strength of the effective Heisenberg coupling and changes
from $\Phi_0$ first to $\Phi_0/2$ and eventually to $\Phi_0/N$
when the ring gets narrower.
\end{abstract}

\maketitle

\section{introduction}

The observation of the Aharonov-Bohm 
oscillations\cite{aharonov1959} and 
persistent current\cite{buttiger1983} 
in small conducting rings, on one hand,
and the recent experimental developments in manufacturing
quantum dots\cite{tarucha1996} and rings\cite{lorke1998}
with only a few electrons, on the other, 
have made quantum rings an ever increasing topic of experimental
research and a new playground for many-particle theory
in quasi-one-dimensional systems.

Many properties of the quantum rings can be explained with 
single-electron theory, which in a strictly one-dimensional
(1D) system is naturally very simple. On the contrary,
the many-particle fermion problem in 1D systems is surprisingly
complicated due to enhanced importance of the Pauli exclusion
principle. In general, correlations are always strong, leading
to non-fermionic quasiparticles as low energy excitations.
It is then customary to say that strictly 1D systems are not  
'Fermi liquids' but 'Luttinger liquids' with 
specific collective excitations (for reviews 
see\cite{haldane1994,voit1994,schulz1995,kolomeisky1996}). 

The many-particle approach normally used in studying the properties
of Luttinger liquids starts by assuming an infinitely long
strictly one-dimensional system. In small finite rings 
a direct diagonalization of the many-body Hamiltonian,
using numerical techniques and a suitable basis,
is possible and can provide more direct information on the 
many-particle states. Two different theoretical models
have been used for the finite rings.
In a {\it discrete model}
the ring is assumed to consist of $L$ discrete lattice
sites (or atoms) with $N$ electrons, which can hop from site to site.
In this case the electron-electron interaction is usually assumed
to be effective only when the electrons are at the same site.
The many-particle 
Hamiltonian is then a Hubbard Hamiltonian\cite{hubbard1963} or its extension.
Another possibility is to assume a ring-shaped smooth
external confinement where interacting electrons move.
In this {\it continuum model} the electron-electron 
interaction is usually the normal Coulomb interaction
($e^2/4\pi\epsilon_0r$). In the case of a small number of 
electrons (typically $N<10$) the many-particle problem is 
well defined in both models and can be solved with  
numerical diagonalization 
techniques for a desired number of lowest energy states.

Other many-particle methods have also been applied for 
studying the electronic structure of quantum rings.
The discrete rings can be solved by using the 
Bethe ansatz\cite{bethe1931,lieb1968},
which becomes powerful especially in the case of an infinitely
strong contact interaction (the so-called $t$-model). 
In the case of continuum rings, quantum Monte Carlo 
methods and density functional methods have been used
(see Sec. \ref{s120}).

The purpose of this paper is to give an introductory review
to the many-particle properties of rings with a few electrons.
Our aim is not to give a comprehensive review of all the
vast literature published. 
The main emphasis is to clarify the relations between 
different methods and to point out general features of
the electronic structures of the rings and their origins.
Most of the results we show in figures are our own computations
made for this review. However, we want to stress that
most of the phenomena shown have been published before,
(in many case by several authors) and we will give reference 
to earlier work.

We take an approach where we analyse the many-body
excitation spectrum and its relation to the single particle spectrum
and electron localization along the ring. We will show that,
irrespective of the model, the excitation spectrum in narrow rings
can be understood as a rotation-vibration spectrum of localized 
electrons. The effect of the magnetic flux penetrating the ring
is also studied as the change of the spectrum as a function of the
flux. This is used to analyze the periodicity of the 
{\it persistent current} as a function of the flux. 
Again, it is shown that similar results are obtained with
the discrete and continuum models.

Throughout this paper we use the term ``spinless electrons'' to
describe a system of completely polarized electron system,
i.e. the many-particle state having maximum total spin and its $z$-component.
We use lower case letters to describe single particle
properties and capital letters for many-particle properties
(e.g. $m$ and $M$ for angular momenta). We use terms like 
``rotational'' and ``vibrational'' states quite loosely for describing
excitations, which in certain limiting cases have exactly those
meanings. When we talk about the rotational state we will use
the terminology of the nuclear physics and call the lowest energy
state of a given angular momentum an {\it yrast} state.

Experimentally, the study of the spectra of quantum
ring is still in an early state of an impressive development.
It is not yet the time to make detailed comparison between the 
theory and experiments. Nevertheless, we will give in Sec. 
\ref{s05} a short overview of the experimental situation.

We then attempt to review the theory of quantum rings in a
logical and pedagogical way, starting with the simplest case
of non-interacting spinless fermions (Secs.\ref{s10} and \ref{freevibsec})
and classical interacting electrons (Sec. \ref{s30}), then introducing
the effect of magnetic flux (Sec. \ref{s40}) and spin (Sec.\ref{s50}).
Lattice models are presented in Secs.\ref{s60} and \ref{s70}, followed
by numerial approaches (Secs. \ref{s90}, \ref{s100},
\ref{s120}). The periodicity properties of the many-body spectrum
is discussed in Sec.\ref{s110}. We also briefly discuss the relation
of these previous approaches to the Luttinger liquid formalism
(Sec. \ref{s130}), and introduce pair correlation functions as a
tool to study the internal structure of the many-electron state
(Sec. \ref{s140}). 
Most of the review will deal with rings where the external magnetic
flux penetrates the ring in such a way that the magnetic field is 
zero at the perimeter of the ring (Aharonov-Bohm flux) and 
the ring is free from impurities. Nevertheless, 
in Sections \ref{s150} and \ref{s160} we will give
short overviews of the effects of the Zeeman splitting and impurities
on the excitation spectrum.

Several interesting aspects of quantum rings will have to be
neglected in this paper. Among these are the exciting possibilities of
observing experimentally the spin Berry phase (see, however, Sec.\ref{s05}
for a brief experimental overview) and, even more exotically,
fractional statistics (anyons) \cite{kane2002}.

\section{Experimental situation}
\label{s05}

Since the mid-eighties, there has been an impressive experimental
development towards smaller and smaller 2D quantum rings; with
the most recent techniques one has reached the true quantum limit 
of nanoscopic rings containing only a few 
electrons\cite{garcia1997,lorke1998,lorke2000}.
At the same time, many of the experiments
still study mesoscopic rings with hundreds of electrons.
Methods of forming such rings include litographic methods 
for forming individual rings on a semiconductor surface. 
The spectroscopic techniques are based on the tunneling current through 
the ring or capacitance spectroscopy. 
Another possiblity is to create a large number of self-organized rings
on a substrate. The large number of rings allows for observation of 
direct optical absorption.

One of the hallmarks in this field of research has been the
experimental observation of the Aharonov-Bohm effect\cite{aharonov1959}
or, equivalently, persistent currents. One of the main challenges
in order to observe this purely quantum mechanical effect, has been to 
ensure phase coherence along the circumference of the ring.
Early experiments in the eighties and nineties reported
observations of Aharonov-Bohm oscillations and persistent
currents in metallic (Au or Cu) rings\cite{webb1985,levy1990,chandrasekar1991}
and in loops in GaAs heterojunctions, i.e. two-dimensional electron
gas\cite{timp1987,ford1989,timp1989,ismail1991,mailly1993,liu1993,liu1994}.
A related effect which has received recent 
theoretical \cite{loss1990,stern1992,aronov1993,qian1994,yi1997} 
and experimental\cite{morpurgo1999,nitta1999,yau2002,yang2002}
attention,
is the occurrence of a spin Berry phase\cite{berry1984} in conducting 
mesoscopic rings. The simplest example of this topological effect
\cite{footnote1}
is the phase picked up by a spin $\frac{1}{2}$ which follows adiabatically 
an inhomogeneous magnetic field; the Berry phase is then proportional 
to the solid angle subtended by the magnetic field it goes through.
It has been shown\cite{aronov1993} that in 1D rings, a Berry
phase may arise due to spin-orbit interactions. The first experimental
evidence of Berry's phase in quantum rings was reported in 1999 by Morpurgo
et al.\cite{morpurgo1999} who interpreted the splitting of certain peaks
in Fourier spectra of AB oscillations as being due to this effect.
Very recently, Yang et al.\cite{yang2002} observed beating patterns
in the Aharonov-Bohm conductance oscillations of singly connected
rings; these results are interpreted as an interference effect
due to the spin Berry phase.

There is by now a vast literature on experiments on quantum rings,
and in the remainder of this section we will just discuss a few
selected papers which report measurements of the many-body
spectra, as this is the main topic of this review.

Fuhrer {\it et al.}\cite{fuhrer2001} used an atomic force microscope 
to oxidize a quantum ring structure on the surface of a AlGaAs-GaAs
heterostructure. By measuring the conductance through the dot they 
could resolve the so-called addition energy spectrum
(see e.g. \cite{reimann2002})
as a function of the magnetic field. The energy levels in the ring
and their oscillation as a function of the magnetic field could 
be explained with a single particle picture assuming small
nonspherical disturbation for the ring. The ring had about 200 
electrons.

Lorke {\it et al.}\cite{lorke1998,pettersson2000,lorke2000} have succeeded
to produce a few electron quantum rings from self-assembled
InAs dots on GaAs using suitable heat treating. They
used far-infrared (FIR) transmission spectroscopy and 
capacitance-voltage spectroscopy to study the ground and 
excited many-body states. The ring radius was estimated to be
$R_0=14$ nm and the confining potential strength $\hbar\omega_0=12$ MeV
(the confining potential is assumed to 
be $\frac{1}{2}m^*\omega_0^2(r-R_0)^2$). Lorke {\it et al.}
were able to study the limit of one and two electrons in the ring.
The experimental findings were in consistence with a single 
electron picture.

Warburton {\it et al.}\cite{warburton2000} studied the photoluminiscence
from self-assembled InAs quantum rings at zero magnetic field. 
By using a confocal microscope
and taking advantage of the fact that each ring had unique charging voltages
they were able to measure the photoluminiscence from a single
quantum ring. The photolumiscence spectra show the effect of the 
Hund's rule of favouring parallel spins and other details of the 
spectra, for example the singlet-triplet splitting. 

The periodicity $\Phi_0=h/e$ of the persistent current 
predicted for normal metal 
(not superconducting) rings has been observed 
several times for semiconductor quantum 
rings\cite{timp1987,timp1989,liu1993,ford1989,ismail1991}.
Also observations of other periodicities, or higher
harmonics, have been reported\cite{liu1994,pedersen2000}
but they have been interpreted as effects of the 
nonperfectness of the rings. 
Very recently, the first observation of the inherent
$\Phi_0/N$ periodicity,
which should appear in perfect very narrow rings,
was reported \cite{keyser2003}.

\section{Strictly 1D-ring with noninteracting spinless electrons}
\label{s10}

The single particle Hamiltonian of an electron in a 
strictly 1D ring depends only on the polar angle $\varphi$
\begin{equation}
H=-\frac{\hbar^2}{2m_eR^2}\frac{\partial^2}{\partial\varphi^2},
\quad \psi_m(\varphi)=e^{im\varphi},
\label{ni1}
\end{equation}
where $R$ is the ring radius, $m_e$ the electron mass
(or effective mass) and $m\hbar$ is the angular momentum. 
(Note that the direction of
the angular momentum axis is always fixed in a 2D structure).\
The corresponding energy eigenvalues are
\begin{equation}
\epsilon_m=\frac{\hbar^2m^2}{2m_eR^2}.
\label{ni2}
\end{equation}

We will first study spinless electrons, or a polarised
electron system, where each electron has a $S_z=\uparrow$.
In the noninteracting case the many-body state is a single 
Slater determinant. The total angular momentum is
\begin{equation}
M=\sum_i^N m_i
\label{ni3}
\end{equation}
and the total energy 
\begin{equation}
E=\sum_i^N \epsilon_{m_i}
\label{ni4}
\end{equation}
The lowest energy state for a given angular momentum,
or the {\it yrast} state, is obtained by occupying single
particle states consecutively (next to each other). This is due to the 
upwards curvature of $\epsilon_m$. We will denote states
consisting of single particle states with
consecutive angular momenta, say from
$m_0$ to $m_0+N-1$, as ``compact states''. These states have energy
\begin{equation}
E_{\rm CS}=\frac{\hbar^2}{2m_eR^2}
[Nm_0^2+(N^2-N)m_0+\frac{1}{6}(2N^3-3N^2+N)]
\label{ni5}
\end{equation}
and the angular momentum 
\begin{equation}
M=Nm_0+\frac{N(N-1)}{2}.
\label{ni6}
\end{equation}
It is interesting to note that while the single particle energy 
increases with the angular momentum as $\hbar^2m^2/2m_eR^2$, 
the lowest many-body energy increases,
in the limit of large $M$, as $\hbar^2M^2/2(Nm_e)R^2$.
An yrast state of the ring with $N$ electrons corresponds then
to a single particle with mass $Nm_e$. This is our first notion
of ``rigid rotation'' of the quantum state.

\begin{figure}
\includegraphics{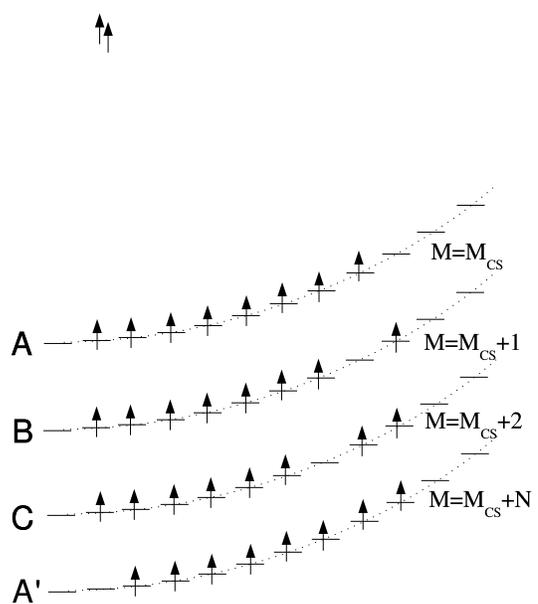}
\caption{
Configurations of the many-body states of eight noninteracting electrons
in a ring for spinless electrons.
A and A' give local minima in the yrast line, while B and C 
correspond to 'vibrational excitations'.
}
\label{f1}
\end{figure}

\begin{figure}
\includegraphics[angle=-90]{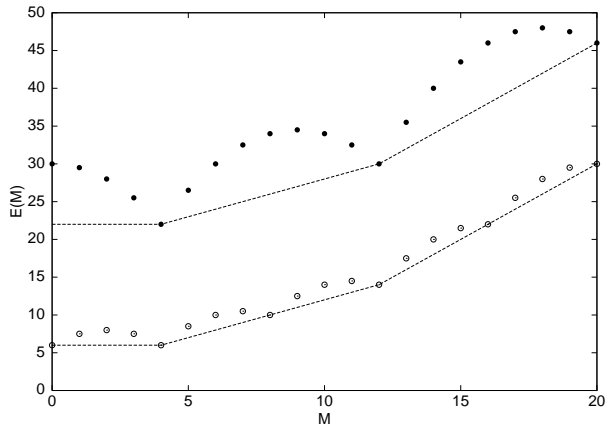}
\caption{
The lowest total energy of eight noninteracting electrons as a function
of the total angular momentum, i.e. the yrast line. 
Black dots: spinless electrons;
open circles: electrons with spin. 
The local minima are connected with dashed lines.
}
\label{f2}
\end{figure}

The structure of the yrast states is illustrated schematically in
Fig. \ref{f1}, and the actual energy levels as a 
function of the angular momentum are shown in
Fig. \ref{f2}. Naturally, the compact states give local minima of the 
yrast spectrum. In what follows, 
the most important property of the many-particle states is
that the {\it internal structure} of the state does not
change when the angular momentum is increased by a multiple of $N$.
(By the term ``internal structure'' of a state, we refer to interparticle
correlations which can be seen by using a rotating frame,
as discussed by Maksym in the case of quantum dots\cite{maksym1996}.)
This follows from the notion that
\begin{equation}
\Psi_{M+\nu N}^{\rm Slater}(\{ \varphi_i \})=\exp \left(i\nu\sum_i^N 
\varphi_i\right)\Psi_{M}^{\rm Slater}(\{ \varphi_i \}),
\label{ni7}
\end{equation}
where $\nu$ is any integer.
This kind of change of the total angular momentum corresponds to 
a rigid rotation of the state and naturally leads to the above mentioned
result that the $N$-electron system rotates like a single 
particle with mass $Nm_e$.
Moreover, correlation funtions are the same for both 
$\Psi_{M+\nu N}^{\rm Slater}$ and $\Psi_M^{\rm Slater}$ since
they are derived from the square $\vert \Psi\vert^2$.
Note that the minima of the yrast spectra occur at angular momenta 
$M=\nu N$ if $N$ is odd and at angular momenta $M=\nu N+N/2$ if
$N$ is even.

\section{Localization of noninteracting spinless fermions in 1D}
\label{freevibsec}
Classical noninteracting particles do not have phase transitions
and are always in a gas phase: There is no correlation between the
particles. In the quantum mechanical case, due to the Pauli 
exclusion principle, one-dimensional
spinless fermions behave very differently: Two electrons can
not be at the same point. This means that noninteracting spinless
fermions are identical with fermions interacting with an infinitely
strong delta-function interaction.
The requirement that the wave function has to go to zero at
points where electrons meet, increases the kinetic energy 
proportional to $1/d^2$ where $d$ is the average distance between 
the electrons (this is analogous to the kinetic energy of a
single particle in a one-dimensional potential box). 
The requirement of the wave function being zero in the contact points
has the classical analogy that the electrons
can not pass each other\cite{schulz1995}.
The pressure of the kinetic energy scales as an interparticle energy
of the form $1/[R^2(\varphi_i-\varphi_j)^2]$, leading to the
interesting result 
that  the energy spectrum of the particles interacting with the 
$\delta$-function interaction agrees with that of particles
with $1/r^2$ interaction. In fact, the model of 1D particles with a  
$1/r^2$ interaction, the so-called Calogero-Sutherland 
model\cite{calogero1969,sutherland1971}, is exactly solvable.
We will return to this in Sec. \ref{s70.3}.

We will now demonstrate that the non-compact states may in fact be
regarded as 
{\it vibrational} states. The simplest case is that
with two electrons. 
The Slater determinant is (omitting normalization)
\begin{equation}
\psi_{m_1m_2}(\varphi_1,\varphi_2)=
e^{im_1\varphi_1}e^{im_2\varphi_2}-e^{im_1\varphi_2}e^{im_2\varphi_1}.
\label{eq201}
\end{equation}
The square of the amplitude of this wave function can be written as
\begin{equation}
\vert \psi_{m_1m_2}(\varphi_1,\varphi_2)\vert^2
=4\sin^2\left[\frac{1}{2}(\Delta m\Delta \varphi)\right],
\label{eq202}
\end{equation}
where $\Delta m=m_1-m_2$ and $\Delta\varphi=\varphi_2-\varphi_1$.
This means that the maxima of this wave function
occur at points $\Delta\varphi=(1+2n)\pi/\Delta m$ where $n$ is an integer.
The compact ground state has $m_1=0$ and $m_2=1$ occupied, i.e.
$\Delta m=1$ while all noncompact states have $\Delta m>1$.
This means that between 0 and $2\pi$ there will be one maximum
for the ground state, the two electrons being  at the opposite
side of the ring. For the 
noncompact states there are two or more maxima between 0 and $2\pi$
and the wave functions will resemble excited states of a 
harmonic oscillator, i.e. vibrational states 
(actually in the two-electron case they are 
exactly those of a single particle in a 1D potential box).

Let us now generalize this analysis to a general case with $N$
electrons. To this end we determine the 
square of the many-body wave function, $\vert\Psi(\{ \varphi_i \})\vert^2$,
in terms of the normal modes of classical harmonic
vibrations. The equilibrium positions of classical
particles are $\varphi_j^0=2\pi j/N$.
The displaced positions of the particles corresponding to
a normal mode $\nu$ are
\begin{equation}
\varphi_i^\nu=\varphi_i^0 + A \sin(\nu(i-\frac{1}{2})2\pi/N),
\label{ni8}
\end{equation}
where $\nu$ is an integer ($1\cdots N/2$) and $A$ the amplitude of
the oscillation.

\begin{figure}
\includegraphics[angle=-90]{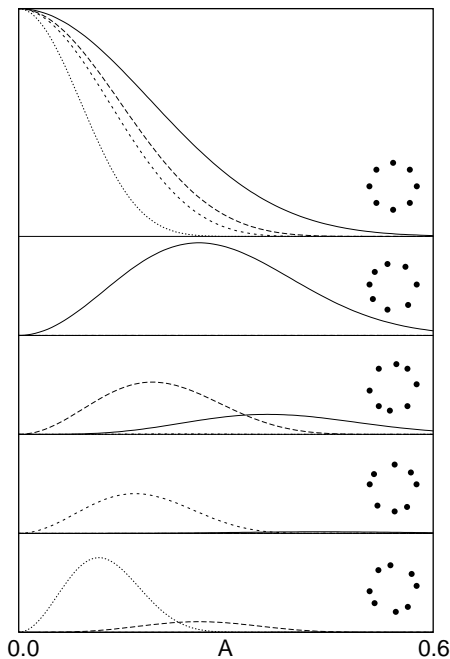}
\caption{
The correlation of different quantum states of noninteracting
electrons with the electron localization at the sites determined
by the classical harmonic vibrational modes,
as a function of the amplitude of the atomic displacement from the
classical equilibrium position,.i.e. 
$\vert\Psi(\{\varphi\})\vert^2$ as a function of $A$ of Eq. (\ref{ni8}).
The panels from top to bottom correspond to
quantum states 
$(\uparrow\uparrow\uparrow\uparrow\uparrow\uparrow\uparrow\uparrow)$, 
$(\uparrow\uparrow\uparrow\uparrow\uparrow\uparrow\uparrow\circ\uparrow)$, 
$(\uparrow\uparrow\uparrow\uparrow\uparrow\uparrow\circ\uparrow\uparrow)$, 
$(\uparrow\uparrow\uparrow\uparrow\uparrow\circ\uparrow\uparrow\uparrow)$, 
$(\uparrow\uparrow\uparrow\uparrow\circ\uparrow\uparrow\uparrow\uparrow)$, 
respectively, where $\circ$ refers to to an empty angular momentum state.
The solid, long-dashed, short-dashed, and dotted lines correspond to
$\nu=1$, 2, 3, and 4, respectively.
}
\label{f3}
\end{figure}

Figure \ref{f3} shows $\vert\Psi(\{ \varphi_i^\nu \})\vert^2$
for the different yrast states of a ring with 8 electrons,
as a function of the amplitude $A$ of the classical oscillation.
In the case of a compact state (denoted by A in Fig. \ref{f1})
$|\Psi|^2$ 
decreases rapidly with increasing $A$ for all $\nu$,
as seen in the uppermost panel. 
This means that particles appear as being localized 
at the sites of classical particles with a repulsive 
interaction.
For the non-compact yrast states (B and C in Fig. \ref{f1} and so on)
the maximum of $\vert \Psi\vert^2$ is reached with a finite 
value of $A$ so that $\nu=1$ corresponds to state B, 
$\nu=2$ corresponds to C, etc. The insets of the figure show the 
positions of the electrons at the maximum $\vert\Psi^2\vert$ of the 
corresponding figure. Clearly, there seems to be
a correspondence between the different yrast states
of {\it noninteracting spinless} electrons
and vibrational modes of classical {\it interacting}
particles in a ring. 

The conclusion of this section is that the yrast spectrum
of noninteracting spinless particles can be understood as 
rotational vibrational spectrum of classical particles 
with a repulsive interaction: The compact states are purely
rotational states, and the non-compact states correspond to
vibrational excitations.

\section{Classical interacting electrons in a strictly 1D ring}
\label{s30}
A finite number of classical interacting particles in a strictly
1D ring will have discrete vibrational frequencies, which, after
quantization, will give the quantum mechanical vibrational energies 
$\hbar\omega_\nu$. We assume a monotonic repulsive pairwise 
potential energy $V(r)$ between the particles, where $r$ is the
direct inter-particle distance. The potential
energy can be written as 
\begin{equation}
E=\frac{1}{2}\sum_{i\neq j}^N 
V\left( 2R\left|\sin\left(\frac{\varphi_i-\varphi_j}{2}\right)\right| \right),
\label{eq30_1}
\end{equation}
where $R$ is the radius of the ring and 
$\varphi_i$ the position (angle) of particle $i$. 
For any pair potential $V$ it is straightforward to solve
(numerically) the vibrational modes. In the case of a short range
potential, reaching only to the nearest neighbours,
one recovers the text-book example of acoustic modes
of an infinite 1D lattice\cite{ashcroft1976} 
(now only discrete wave vectors are allowed
due to the finite length $2\pi R$)
\begin{equation}
\hbar\omega_\nu=C \sin\left(\frac{\nu \pi}{N} \right),
\label{eq30_2}
\end{equation}
where $C$ is a contant (proportional to the velocity of sound)
and $\nu$ an integer.

\begin{figure}
\includegraphics[angle=-90]{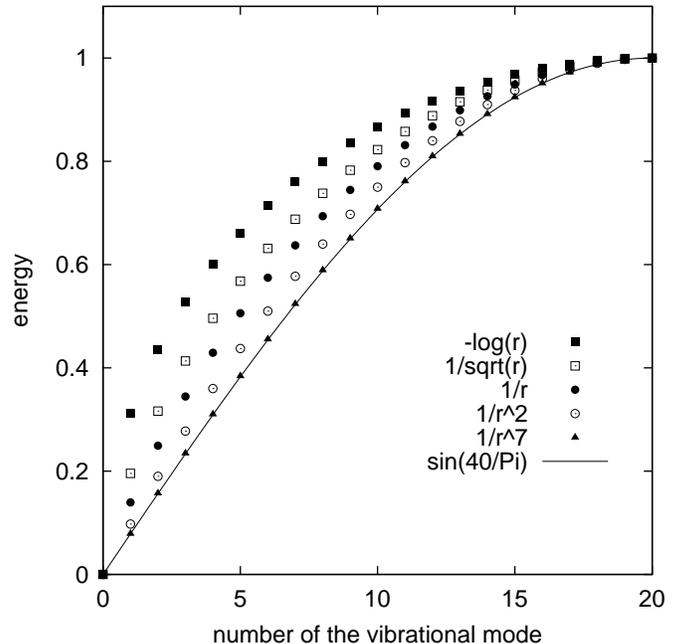}
\caption{Vibrational modes of 40 classical particles in a 1D ring.
Different symbols correspond to different forms of the repulsive
interaction between the particles, as indicated in the figure.
The solid line is the result of a nearest neighbour harmonic model.
The energies are scaled so that the highest vibrational energy for
each potential is one.}
\label{f4}
\end{figure}

Figure \ref{f4} shows the classical vibrational energies for
different pair potentials $-\ln(r)$, $1/\sqrt{r}$, $1/r$, $1/r^2$,
and $1/r^7$. Naturally, when the range of the potential gets shorter
the vibrational energies approach that of the nearest neighbour
interaction, Eq.(\ref{eq30_2}), also shown. Note that the vibrational energies
of the Coulomb interaction ($1/r$) do not differ markedly from
those of $1/r^2$-interaction. The latter has the special property
that the energies agree exactly with the quantum mechanical energies of 
noninteracting spinless fermions, as discussed in section \ref{s70.3}.

In addition to the vibrational energy, the classical system can have 
rotational energy determined by the angular momentum:
$E_{\rm rot}=\frac{1}{2}NmR^2\dot\theta^2$
($NmR^2$ being the moment of inertia and 
$\dot\theta$ the angular velocity).
Quantization of the vibrational and rotational energies
gives the energy spectrum
\begin{equation}
E=E_{\rm rot}(M)+E_{\rm vib}(\{n_\nu\})=\frac{\hbar^2M^2}{2NmR^2}
+\sum_\nu n_\nu\hbar\omega_\nu,
\label{e30_3}
\end{equation}
where $M$ is the (total) angular momentum and 
$n_\nu$ the number of phonons $\nu$.

\begin{figure}
\includegraphics[angle=-90]{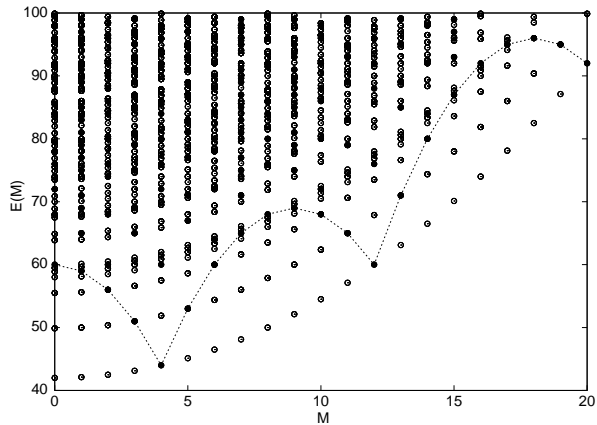}
\caption{Low-energy spectrum of eight electrons interacting with
infinitely strong $\delta$-function interaction.
The black dots indicate the state with maximum spin
('spinless electrons'). White circles give the energies of other spin states.
The spectrum is identical with the rotation-vibration spectrum of 
eight particles interacting with $1/r^2$-interaction.
}
\label{f5}
\end{figure}

Figure \ref{f5} shows the rotational vibrational spectrum derived
from Eq.(\ref{e30_3}) for eight particles with $1/r^2$ interaction. 
It agrees exactly with that calculated quantum mechanically
for electrons with (an infinite) $\delta$-function interaction
(see Section \ref{s70}). It is important to notice that since the 
vibrational states are fairly independent of the interparticle
interaction, the quantum mechanical spectrum close to the yrast
line is expected to be be qualitatively the same 
irrespective of the interaction.

Quantum mechanics plays an important role, however, when the 
Pauli exclusion principle is considered. The requirement of the 
antisymmetry of the total wave function restricts what spin-assignments
can be combined with a certain rotation-vibration state.
For example, for a completely polarized ring (maximum spin)
only certain rotational vibrational states are allowed, as shown
in Fig. \ref{f5}. Group theory can be used to analyze the possible 
spin-assignments\cite{koskinen2001,koskinenp2002} 
in a similar way as done in rotating 
molecules\cite{herzberg1945,tinkham1964}.

In the case of bosons (with $S=0$) the total wave function has to
be symmetric, and the allowed rotation-vibration states are
exactly the same as for spin-1/2-particles in the fully polarized
state. Figure \ref{f5} is then also a 'general' result for eight bosons
in a quantum ring (assuming the interaction to be repulsive).

\section{Effect of a magnetic flux -- persistent currents}
\label{s40}

We consider a magnetic flux going through the quantum ring
in such a way that the magnetic field is zero at the radius
of the ring. This can be modelled, for example, by 
choosing the vector potential to be (in circular cylindrical
coordinates)
\begin{equation}
A_r=A_z=0,\qquad
A_\varphi=\begin{cases}\frac{B_0r}{2},\textrm{if } r\leq r_c\\ 
\frac{B_0r_c^2}{2r}=\frac{\Phi}{2\pi r},\textrm{if } r>r_c
\end{cases}
\label{noa1}
\end{equation}
which gives a flux $\Phi=\pi r_c^2B_0$ penetrating the ring in such
a way that the field is constant inside $r_c$ and zero outside.
If the ring is in the field-free region ($R>r_c$) the electron
states depend only on the total flux penetrating the ring.

The solutions of the single particle Schr\"odinger equation of 
an infinitely narrow ring ,
\begin{equation}
\frac{1}{2m_e}\left(-\frac{i\hbar}{R}\frac{\partial}{\partial\varphi}
- \frac{e\Phi}{2\pi R}\right)^2\psi(\varphi)=\epsilon\psi(\varphi),
\label{noa2}
\end{equation}
can be still written as flux-independent 
plane waves $\psi_m = \exp(im\varphi)$, but the 
corresponding single particle energies now depend on the flux as 
\begin{equation}
\epsilon(m,\Phi)=\frac{\hbar^2}{2m_eR^2}\left(m-\frac{\Phi}{\Phi_0}\right)^2.
\label{moa3}
\end{equation}
The many-body wave function is still the same Slater determinant as
without the field, 
while the total energy becomes
\begin{equation}
E(M,\Phi)=E(M,0)-\frac{\hbar^2M}{m_eR^2}\left(\frac{\Phi}{\Phi_0}\right)+
\frac{\hbar^2N}{2m_eR^2}\left(\frac{\Phi}{\Phi_0}\right)^2
\label{ni9}
\end{equation}
The effect of increasing flux is not to change the 
level structure but to tilt the spectrum, say of Fig. \ref{f2},
such that the global minimum of the total energy jumps
from one compact state to the next compact state.
Note that Eq.(\ref{ni9}) is true also for interacting electrons in
a strictly one-dimensional ring. 
This follows from the fact that any good
angular momentum state can be written as a linear combination of 
Slater determinants of non-interacting states. For each of these,
the last two terms are the same, while $E(M,0)$ depends on the
interactions.

Alternatively, one may perform a unitary transformation to obtain a
description of the system in terms of a {\em field-free} Hamiltonian, but
with multivalued wavefunctions (``twisted boundary conditions'').
Let us choose a gauge ${\bf A}=\nabla \chi$ (which
ensures that ${\bf B}=\nabla\times {\bf A}$ is zero at the ring;
in a strictly one-dimensional ring one may write 
$\chi(\varphi) = \Phi\, \varphi/(2\pi)$
and consider the unitary transformation
\be{unit1}
\psi \rightarrow \psi' &=& U \psi \\
H \rightarrow H' &=& UHU^{-1}
\ee
where the operator $U$ is defined as
\begin{equation}
U = e^{-ie/\hbar \int {\bf A}\cdot d{\bf l} } = e^{-ie/\hbar \chi}.
\label{unit2}
\end{equation}
Obviously, the eigenspectrum will be conserved under this transformation 
-- if $H\psi = E\psi$, then $H'\psi' = UHU^{-1}U\psi = UH\psi =E\psi'$.
It is easy to show that the effect of this transformation on the Hamiltonian 
is to cancel out the gauge field, i.e.
\be{unit3}
U(\frac{-i\hbar}{R} \frac{\partial}{\partial\varphi})U^{-1}
= \frac{-i\hbar}{R} \frac{\partial}{\partial\varphi} + e A_{\varphi},
\ee
and the Hamiltonian takes the field-free form
\begin{equation}
H' = -\frac{\hbar^2}{2m_eR^2}\frac{\partial^2}{\partial\varphi^2}.
\label{noa7}
\end{equation}
Meanwhile, the wave function now picks up a phase $(-ie/\hbar)\chi$ 
when moving along a given path, even though it moves in a region
where the magnetic field is zero. This is the so-called Aharonov-Bohm
effect\cite{aharonov1959}. In particular, if an electron moves along a 
closed path around the flux tube, the Aharonov-Bohm phase becomes
\begin{widetext}
\begin{equation}
\frac{-ie}{\hbar}\Delta\chi
=\oint_{2\pi R} \nabla\chi\cdot d{\bf l}
=\frac{-ie}{\hbar}\oint_{2\pi R}{\bf A}\cdot d{\bf l} 
=\frac{-ie}{\hbar}\int_{\pi R^2} {\bf B}\cdot d{\bf s}
=\frac{-ie}{\hbar}\Phi.
\label{noa5}
\end{equation}
\end{widetext}
In other words, the boundary condition has now changed. 
While the original wave function satisfies periodic boundary
conditions, $\psi(\varphi)=\psi(\varphi+2\pi)$, for the new 
wave function we have the condition
$\psi'(\varphi+2\pi)=\psi'(\varphi)\exp(-ie\Phi/\hbar)$,
i.e. ``twisted boundary conditions''.
Note that this boundary condition naturally leads to periodic 
eigenvalues
$\epsilon'(m')=\epsilon'(m'+\Phi/\Phi_0)$, in the same way as the 
Bloch condition for electron states in a periodic
lattice leads to the periodicity of
the eigenenergies in the reciprocal lattice\cite{ashcroft1976}.

%
The spectrum (\ref{moa3}) is clearly periodic in flux with period
$\Phi_0$. Any {\em given} eigenstate 
$\psi'_m = \exp(i(m-\Phi/\Phi_0)\varphi)$, however,
will have its angular momentum eigenvalue shifted by one as the flux
is changed by one flux quantum.

The persistent current of a quantum ring can be written in
general as 
\begin{equation} 
I(\Phi)= -\frac{\partial F}{\partial \Phi},
\label{ni10}
\end{equation}
where $F$ is the free energy of system.
To illustrate this in the simplest possible case, consider
the Schr\"odinger equation for a one-electron ring,
\be{curr1}
-\frac{\hbar^2}{2m} D^2 \psi_m(\varphi) = E_m \psi_m(\varphi)
\ee
where 
\be{curr2}
D = \frac 1 R \frac{\partial}{\partial \varphi} 
	- \frac{ie}{\hbar} A_{\varphi}.
\ee
Multiplying both sides of Eq.(\ref{curr1}) by $\psi_m^*$
from the left and integrating along the circumference of the ring
one gets
\be{curr3}
E_m = -\frac{1}{2m_e} \int_0^{2\pi} R d\varphi \psi_m^*(\varphi) D^2
	\psi_m(\varphi).
\ee
Using the expression (\ref{noa1}) for the gauge field and taking the
derivative with respect to flux one obtains, after an integration
by parts,
\begin{widetext}
\begin{equation}
\frac{\partial E_m}{\partial \Phi}
 = \frac{1}{2\pi R}\, \frac{ie\hbar}{2m_e} \int_0^{2\pi}
   R d\varphi \left[ \psi_m^* D \psi_m - \psi_m D^* \Psi_m^*\right] 
 = - \frac{1}{2\pi} \int_0^{2\pi} d\varphi \,j(\varphi)
 \equiv -I_m,
\label{curr4}
\end{equation}
\end{widetext}
where we have identified the RHS with the 1D current (density)
associated with the state $m$. (Note that $j(\varphi)$ has to
be independent of the angle $\varphi$). Obviously, the same
argument applies for a non-interacting many-body system where
the contributions from different angular momentum states are
simply summed. The same is true in the presence of interactions
due to the fact that all flux dependence is in the kinetic
energy term of the many-body Hamiltonian 
(see discussion after Eq.(\ref{ni9})).

Due to the periodicity of the energy spectrum, the persistent 
current will be a periodic function of the flux. 
In the case of noninteracting spinless electrons, the
period is $\Phi_0$, owing to the fact that the minimum
energy for any flux corresponds to a compact state.
As will be seen in Sec. \ref{s110}, the electron-electron
interactions can change the periodicity to
$\Phi_0/2$ or to $\Phi_0/N$.

\section{Noninteracting particles with spin}
\label{s50}

\begin{figure}
\includegraphics{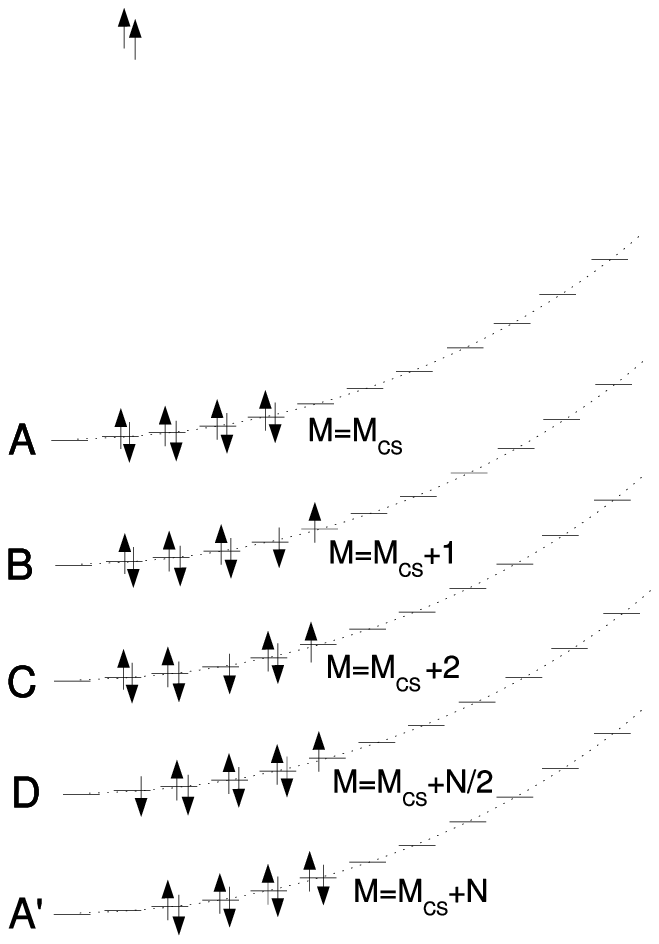}
\caption{
Configurations of the many-body states of eight noninteracting electrons
with spin in a ring.
A, A', and D give local minima in the yrast line, while B and C 
correspond to 'vibrational excitations'.
}
\label{f6}
\end{figure}

The spin degree of freedom allows two electrons for 
each single particle orbital. The yrast states 
of the many-body spectrum are still consisting of
compact or nearly compact states, but now for each 
spin component as shown in Fig. \ref{f6}.
The corresponding yrast spectrum for 8 electrons is shown in Fig. \ref{f2}
in comparison to the spectrum of spinless electrons.
The total spin of the states A (in Fig. \ref{f6}) 
is $S=0$ while for all other states have either 
$S=0$ (singlet) or $S=1$ (triplet) as can easily be
deducted from Fig. \ref{f6} by considering the possible ways
to arrange the $S_z$-components in the orbitals with
only one electron.

The yrast spectrum now consists of downward cusps 
at angular momenta $M=nN/2$, but those minima corresponding to
$S=0$ compact states are deeper. In the case of noninteracting
electrons with spin there is not such a clear relation to 
classical rotation-vibration spectrum as in the case of
spinless noninteracting case discussed above. One could 
consider spin-up and spin-down states separately as 
spinless systems which do not interact with each other.
However, considering the complete many-body spectrum one should
notice that the energy of state B in Fig. \ref{f6} is degenerate
with the state where the spin-up system has five electrons and the
spin-down system only three.

The persistent current of the noninteracting 
system with the spin is again a periodic function of
the flux. The period is $\Phi_0$ like in the case of 
spinless electrons. When the field is increased, the 
ground state shifts from a compact state (of the kind A in Fig. \ref{f6})
to the next similar state A', i.e. the angular momentum shifts with $N$.
At the transition point the three levels, A, D and A',
are actually degenerate, but the D-states do not
change the periodicity of the persistent current.
However, the amplitude of the persistent
current is only half of that of the spinless electrons.

So far we have considered only even numbers of electrons. 
In the case of an odd number 
of electrons, the spinless case is very similar to that of 
even numbers,
the only difference being a phase shift of $\Phi_0/2$ of the 
periodic oscillations.
However, in the case of electrons with spin, the odd number 
of electrons has an
effect also on the amplitude of the persistent current: For a large number
of electrons the amplitude for odd numbers of electrons is exactly 
twice that of 
even numbers of electrons, as first shown by Loss and Goldbart\cite{loss1991}.
This effect is sometimes referred to as a parity effect.

\section{Noninteracting electrons in a lattice}
\label{s60}

Instead of a continuum ring, we will now consider
non-interacting electrons in a 
strictly 1D ring with a strong periodic potential. 
In this case the standard solid state physics approach is
the tight-binding model where the Hamiltonian matrix can
be written as 
\begin{equation}
H_{ij}=\begin{cases}
\epsilon_0 & {\rm if} \quad  j=i \\
-t & {\rm if} \quad j=i\pm 1, \\
\end{cases}
\label{tb1}
\end{equation}
where the diagonal terms describe the energy level 
at a lattice site and the off-diagonal terms describe 
the hopping between the levels. This simplest 
tight-binding model assumes one bound state per lattice site
and is often called the H\"uckel model or CNDO-model
(complete neglect of differential overlap).

Assuming the ring, with radius $R$, to have $L$ lattice sites,
the problem of solving the eigenvalues becomes a text-book
example of 1D band structure\cite{ashcroft1976}, and
the single electron eigenvalues can be written as
\begin{equation}
\epsilon(k)=\epsilon_0-2t\cos(ka)=\epsilon_0-2t\cos\left(\frac{2\pi}{L}m\right),
\label{tb2}
\end{equation}
where $a$ is the lattice constant, $k$ the wave
vector and $m$ an integer. The last step follows from 
the facts that the lattice constant is $a=2\pi R/L$, and in a 
finite ring $k$ will have only discrete values $k=m/R$.
Notice that in the large $L$ limit, $m/L\rightarrow 0$, 
the energy spectrum is equivalent with that of free electrons,
Eq. (\ref{ni2}), if we choose the tight-binding parameters as 
$\epsilon_0=2t$ and $t=\hbar^2/2m_ea^2$. In this case 
the quantum number $m$ gets the meaning of the orbital angular momentum.
(This equivalence of the tight-binding model and free
electron model is valid for simple lattices of any dimension and 
can be derived also by discretizing the Laplace operator of 
the free particle Schr\"odinger equation\cite{manninen1991}).

The many-body state is again a simple Slater determinant of the 
single particle states with the total energy
\begin{equation}
E=N\epsilon_0-2t\sum_{j=1}^N \cos\left(\frac{2\pi}{L}m_j\right),
\label{tb3}
\end{equation}
where the selection of the 'angular momenta' $m_j$ is 
restricted by the Pauli exclusion principle.
For example, in the polarized case (spinless electrons) all $m_j$
must be different, and the ground state is obtained with
a compact state where the $m_j$'s are consecutive integers as
in the case of free electrons. Similarly, we can identify
'vibrational states' by making a hole in the compact state
as will be demonstrated in more detail in the following sections.

\section{Interacting electrons on a lattice: The Hubbard model}
\label{s70}
\subsection{Model and exact diagonalization}
A much studied approach to quantum rings and persistent currents
is the Hubbard 
model\cite{fye1991,schofield1991,kusmartsev1991,yu1992,stafford1993,kusmartsev1995,berciu2000,eckle2001}. 
It describes electrons on a
discrete lattice with the freedom to hop between lattice sites,
and the Coulomb interaction is represented by an on-site repulsion.
An interesting feature of the Hubbard ring is that, despite
being a strongly correlated electron system, it can be solved
exactly. 
For a small number of particles the solution can be found by 
direct diagonalization of the Hamiltonian matrix as discussed
first. For any number of particles another
solution technique, the Bethe ansatz, can be used. This 
technique is most suitable in the limit of infinite $U$
(so-called $t$-model) and will be
addressed in the next subsection. 
The Hamiltonian describing the Hubbard model for an $N$-electron 
ring with $L$ lattice sites, in the presence of a magnetic flux 
$\phi=\Phi/\Phi_0$ piercing the ring, can be derived with the help
of the unitary transformation introduced in Sec.\ref{s40}
(the flux dependence was first derived by Peierls\cite{peierls1933}):
\begin{widetext}
\be{eq:hub}
H = -t \sum_{i=1}^N \sum_{\sigma} 
  \left( e^{-i2\pi\phi/L} c_{i+1, \sigma}^{\dagger} c_{i, \sigma} 
  +  e^{i2\pi\phi/L} c_{i, \sigma}^{\dagger} c_{i+1, \sigma} \right)
 + U \sum_{i=1}^N \hat n_{i \uparrow} \hat n_{i\downarrow}.
\ee
\end{widetext}
Here, the operator $c_{i, \sigma}^{\dagger}$ ($c_{i, \sigma}$)
creates (annihilates) an electron with spin $\sigma$ at site $i$; 
$\hat n_{i\sigma} = c_{i, \sigma}^{\dagger} c_{i, \sigma}$ is
the number operator for spin-$\sigma$ electrons at site $i$. The first
part of (\ref{eq:hub}) describes the hopping of electrons
between neighbouring sites (``kinetic term'') while the last
part gives the repulsion between electrons occupying the same
site. We will set the hopping parameter $t=1$ for simplicity.

For small numbers of electrons $N$ and lattice sites $L$ the Hubbard
Hamiltonian can be solved exactly by diagonalization of the Hamiltonian
matrix. We use an occupation number basis (see e.g. \cite{fetter1971})
$\vert \Psi_\alpha\rangle=\vert n_{\alpha 1\uparrow},n_{\alpha 2\uparrow} \cdots 
n_{\alpha L\uparrow};n_{\alpha 1\downarrow}\cdots n_{\alpha L\uparrow}\rangle$
where the $z$-component of the total spin is fixed, i.e.
$\sum_i n_{\alpha i\uparrow}=N_\uparrow$ and 
$\sum_i n_{\alpha i\downarrow}=N_\downarrow$ ($N=N_\uparrow+N_\downarrow$).
Taking matrix elements $\langle \Psi_\alpha\vert H\vert \Psi_{\alpha '}\rangle$
gives us a matrix with dimension ${L\choose N_\uparrow}{ L\choose N_\downarrow}$.
The eigenvalues of this matrix are the exact many-body energy levels of the 
Hubbard Hamiltonian. 
For a given total spin $S$, the energy eigenvalues do not depend
on $S_z$ (there is no Zeeman splitting since we
assume the magnetic field to be nonzero only inside the ring).
In the case of an even $N$ we can
choose $S_z=0$, or $N_\uparrow=N_\downarrow=N/2$, and the diagonalization of
the Hamiltonian will give us all possible eigenvalues
(for odd $N$ we take $S_z=1/2$).
The easiest way to solve the {\it total} spin of a given 
energy state is then to repeat the computation with all possible values of
$S_z$ and look at the degeneracies. 
(Note that the matrix dimension is largest for $S_z=0$ and thus
solving the matrix for all $S_z>0$ takes less computation than soving the
$S_z=0$ case).

\subsection{The $t$-model}
\label{sec:tm}
In the following we will first focus on the strong repulsion limit 
$U\rightarrow \infty$,
which was first discussed nearly 20 years ago\cite{beni1973}; we shall return 
to finite $U$ effects later.
In this limit, the system is equivalently described by the simpler
``$t$-model'' Hamiltonian\cite{bernasconi1975,vollhardt1994}
\be{eq:tmodel}
H_t = P H_{kin} P,
\ee
where $H_{kin}$ is the kinetic (hopping) term of (\ref{eq:hub})
and $P$ denotes a projection operator which eliminates all
states with doubly occupied sites. It can be shown 
\cite{bernasconi1975} that this projected  Hamiltonian is equivalent
to a tight-binding Hamiltonian describing {\it spinless fermions}.
Moreover, going to large but {\em finite} $U$, near half filling, 
the Hubbard model reduces to the so-called $t-J$ model, i.e. the $t$-model
(\ref{eq:tmodel}) plus a Heisenberg term 
$J\sum_i {\bf S}_i\cdot{\bf S}_{i+1}$ \cite{vollhardt1994}. 
In other words the translational and spin degrees of freedom get decoupled.
We will return to the $t-J$ model in Sec. \ref{s80}.

In the next subsection we will describe how
solutions of the $t$-model can be constructed using the Bethe ansatz.
However, for a small number of electrons and
lattice sites ($N$ and $L$) 
the direct diagonalization has the advantage 
of giving all the eigenvalues at once and some information
of the many-body state is more transparent. The results 
shown for small discrete rings are obtained either 
with diagonalization techniques or using the Bethe ansatz
solution. It should be stressed that both methods are exact and
give the same results.

\begin{figure}
\includegraphics[angle=-90]{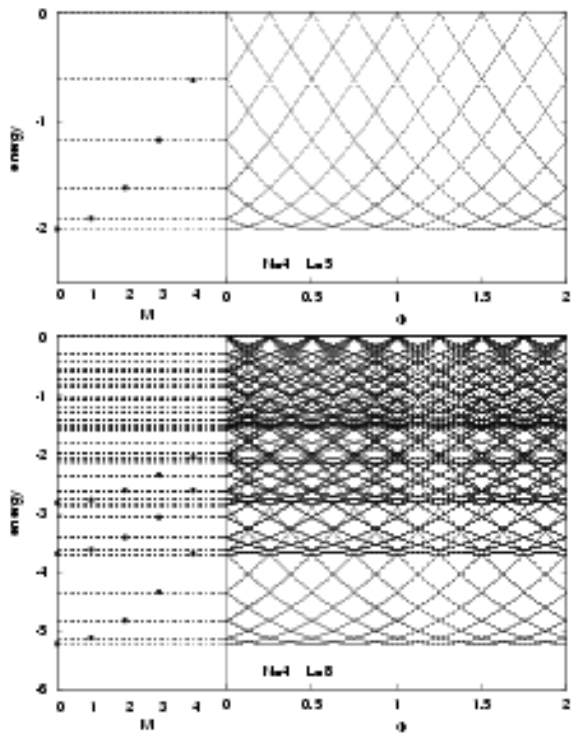}
\caption{Energy spectra of rings of four electrons in five
(upper panel) and eight (lower panel) sites, calculated with 
the $t$-model Hamiltonian. The right hand side shows the 
energy levels as a function of the flux (in units of $\Phi_0$).
On the left, the zero flux energy levels are shown as a function of 
the angular momentum $M$ (black dots). Note that the $L=8$ case
shows several vibrational bands. Only negative energy levels are
shown since the spectrum is symmetric around zero energy.
}
\label{f7}
\end{figure}

We will first study four electrons (with spin) in a lattice having
at least one empty site. Figure \ref{f7} shows the energy
spectrum for $L=5$ and 8. The right-hand panels show the energies
as a function of the magnetic flux piercing the ring, while the 
left-hand side shows the lowest energies at zero flux,
analyzed with respect to the total angular momentum, for the
different ``vibrational bands'' in the spectra.
The following general features should be noted: 
The energy (and thus the persistent current) has a periodicity 
$\Phi_0/N$. In particular,
as first discussed by Kusmartsev\cite{kusmartsev1991} 
and by Yu and Fowler\cite{yu1992},
in the ground state there are
$N$ cusps in every flux period; increasing the flux makes the ground 
state jump from one angular momentum state to the next. 
(A more general discussion of the periodicity of the ground state
will be given in section \ref{s110}.)
Each {\it individual} state, on the other
hand, is a harmonic (cosine) function of the flux, with periodicity
$L\Phi_0$, in accordance with the periodicity of the Hamiltonian.
Thus, the energy of each individual state has the form, up to an
overall phase shift,
\be{eq:stateen}
E \sim \cos\left[\frac{2\pi}{L}\left( -\frac{\Phi}{\Phi_0} 
                  -\frac{p}{N} \right)  \right]
\ee
with $p$ an integer.
Obviously, the (kinetic) energy collapses to zero in the case of half 
filling, $L=N$, as in this case there is no freedom to hop. 
Surprisingly, however, the above generic flux dependence
of the lowest states of the spectra
is obtained even for a single free site (i.e. $L=5$ for $N=4$).
For $L > N+1$, there are several ``energy bands'' consisting of
states with different amplitude; the number of bands increases with
the number of empty sites. As will be discussed below (Sec. \ref{s70.2}), 
the lowest energy state at any given flux corresponds to the rotational 
band without vibrational energy, while the higher bands can be interpreted 
as the vibrational states of the system.

It should be noted that each individual energy level in Fig. \ref{f7}
still has a spin degeneracy: Two or more states with different 
{\it total} spin $S$ belong to each energy level. The magnetic field can
not separate these since we have neglected the possible Zeeman splitting.
A finite $U$ will separate the states belonging to different total spin
as will be shown in Sec. \ref{s80}, but will still leave the degeneracy due to 
the $S_z$.

\subsection{Bethe ansatz}
\label{s70.b}

The Hubbard model in 1D can in fact be solved exactly, in terms
of the Bethe ansatz, as shown by Lieb and Wu in 1968\cite{lieb1968}.
The corresponding solution for a Hubbard ring in the presence 
of an Aharonov-Bohm flux has later been discussed by a number of authors
\cite{shastry1990,kusmartsev1991,yu1992,kusmartsev1995,kotlyar1998}.
Bethe ansatz solutions can be used
to construct not only the ground state but in fact the entire spectrum
presented in the previous section. According to this method,
the energy of a given
many-body state may be written as
\be{eq:BA1}
E = -2 \sum_{j=1}^N \cos \, k_j
\ee 
where the numbers $k_j$ 
can be found by solving the set of Bethe equations
\begin{widetext}
\be{eq:bethe1}
Lk_j=2\pi I_j-\Phi-\sum_{\beta=1}^{N_\uparrow} 2\tan^{-1}[4(\sin k_j-\lambda_\beta)/U] \\
\sum_{j=1}^N 2\tan^{-1}[4(\lambda_\alpha -\sin k_j)/U]=
2\pi J_\alpha+\sum_{\beta=1}^{N_\uparrow} 2\tan^{-1}[2(\lambda_\alpha -\lambda_\beta)/U].
\ee
\end{widetext}
For obtaining the whole energy spectrum, the unknown (complex) constants 
$k_j$ and $\lambda_\beta$ have to be solved for different quantum
numbers $I_j$ and $J_{\alpha}$, which are 
restricted to be integers or half-integers
depending on the numbers $N$ and $N_\uparrow$. For example, if $N$ and
$N_\uparrow$ both are even, $I_j$ must be an integer and $J_j$ a half-integer.
The total (canonical)
angular momentum corresponding to the quantum numbers is given
by
\be{eq:M}
M=\sum_{j=1}^N I_j+\sum_{\alpha=1}^{N_\uparrow} J_\alpha.
\ee
In the general case of a finite $U$ the nonlinear 
Bethe equations turn out to be difficult to solve numerically, at least
for some quantum numbers\cite{karbach1997}. 
For small systems ($N$ and $L$ small)
it is easier to find {\it all} eigenvalues by a direct diagonalization of 
the Hamiltonian matrix as explained in the previous section.

In the limit of infinite $U$, however, the Bethe ansatz solution
becomes particularly simple. In this case
the quantum numbers $k_j \, \epsilon \, [0,2\pi\rangle$ 
are simply given by\cite{kusmartsev1991,yu1992}
\be{eq:BA2}
k_j = \tpl \left[ I_j -\frac{p}{N} + -\pn \right]
\ee
with 
\be{eq:BA3}
p = -\sum_{\alpha=1}^{N_\uparrow} J_\alpha.
\ee
So a given solution
is constructed for a fixed $N_\uparrow$, 
i.e. a fixed value of the $z$-component
of the spin. The quantum numbers
$I_j$ and $J_{\alpha}$ are related to the charge- and spin degrees
of freedom, respectively. In the case we shall consider, even $N$,
the $I_j$:s are integers (half-odd integers) and the $J_{\alpha}$:s 
are half-odd integers (integers) if $N_\uparrow$ is even (odd). 
In practice, all eigenvalues can be found by letting $p$ run over
all integers and choosing all possible sets of $I_j$:s with the restriction
$I_{max} - I_{min} < L$.

\noindent
{\it Ground state of the $t$-model} \\
In analogy to the non-interacting case, the states forming the outermost 
band (see Fig. \ref{f7}), i.e. states which become the ground state at
some value of the flux, are compact states in the quantum numbers  $\{ I_j \}$.
At zero flux, these are of course just the yrast states (see left-hand panels in
Fig. \ref{f7}).
For example, if $N$ and $M$ are both even, the quantum numbers $\{ I_j \}$ 
are consecutive
integers, $I_j = -N/2, -N/2 + 1, ..., N/2-1$. The corresponding
total energy, Eq.(\ref{eq:BA1}), as function of flux, becomes\cite{yu1992}
\be{eq:BA4}
E_0 = -E_m \cos\left[ \tpl\left( -\pn - \frac{p}{N} + D_c \right) \right],
\ee
where $D_c = (I_{max} + I_{min})/2$ and
\be{eq:BA5}
E_m = 2 \, \frac{\sin(N\pi/L)}{\sin(\pi/L)}.
\ee
The integer $p$ should be chosen, for given flux, such as to
minimize the energy. Of course, there are in general several
ways of choosing the ``spin quantum numbers'' $J_{\alpha}$ to give the same sum,
leading to a large degeneracy of the state. This has to do with the fact 
that spin excitations are massless in the limit of the $t$-model; for
finite $U$ this degeneracy would be lifted. 
Note that the energy collapses to zero for half filling, i.e. when $L=N$.

As a simple illustration, consider the example $N=4$ at zero
flux. Then, $I_j = -2, -1, 0, 1$, $D_c = -1/2$ and 
\begin{widetext}
\be{eq:BA6}
\sum_j \cos k_j 
  &=& \cos\left[ \tpl \left( -\frac p N - \frac{1}{2} - \frac{3}{2} \right) \right]
    + \cos\left[ \tpl \left( -\frac p N - \frac{1}{2} - \frac{1}{2} \right)  \right]
\\
  &+& \cos\left[ \tpl \left(  -\frac p N - \frac{1}{2} + \frac{1}{2} \right) \right]
    + \cos\left[ \tpl \left(  -\frac p N - \frac{1}{2} + \frac{3}{2} \right) \right]
\\
   &=& 2 \left(\cos \frac{\pi}{L} + \cos \frac{3\pi}{L}  \right)
         \cos \left[ \tpl \left( \frac p N+ \frac{1}{2}\right) \right] 
\\
   &=& \frac{\sin(4\pi/L)}{\sin(\pi/L)}
         \cos \left[ \tpl \left(  \frac p N + \frac{1}{2}\right) \right].
\ee
\end{widetext}
It is easy to check that the ``amplitudes'',
$-2 \sin(4\pi/L) / \sin(\pi/L)$ agree with those in Fig. \ref{f7}
for $L = 5$ and $8$.
Note that the choice of the ``central value'' $D_c$ as an overall
phase shift makes the sum symmetric; one obtains {\it pairs},
$\cos \left[ \tpl \left( -p/N + D_c \pm \gamma\right) \right]$, so
that, when writing out the sum, all $\sin(2\pi\gamma/L)$-terms cancel.
This also implies that shifting all the $I_j$:s by the same
amount, does not alter the total energy (\ref{eq:BA4}); it
only changes the overall phase shift $D_c$.
The integer $p$ is related to the angular momentum $M$,
see Eq.(\ref{eq:M}).
This construction gives not only the ground state but, for fixed
flux, all energy levels of the lowest band, corresponding to
different $p$. 

\noindent
{\it Higher bands} \\
Generally, excitations can be constructed within the Bethe ansatz 
by introducing {\it holes} in the ground state distribution of 
the quantum numbers $I_j$ and $J_{\alpha}$\cite{kotlyar1998}.
Naturally, these states are related to the ``non-compact'' states
of noninteracting fermions in a strictly 1D ring, discussed in Sec. \ref{s10}.
The higher bands in our spectra correspond to charge excitations,
i.e. holes in the $I_j$ ($J_{\alpha}$-excitations do not cost any energy
in the infinite $U$ limit), and this reproduces exactly all the
energies obtained numerically in the previous section. We will
illustrate this procedure with a few examples for $N=4$, $M=2$.
The lowest possible ``excitation'' in the charge quantum numbers
$I_j$ is lifting the topmost one by one step,
\be{eq:band1}
\{ I_j \} = -2, -1, 0, 2,
\ee 
and the corresponding $k_j$ as given in Eq.(\ref{eq:BA2}). 
The resulting energy is
\begin{widetext}
\be{eq:band2}
E_1 &=& -2 \sum_j \cos k_j \\
  &=& -2 \left(1 + 2\cos \frac{4\pi}{L} + \cos \frac{2\pi}{L}  \right)
         \cos \left[ \tpl \left( -\frac p N - \pn\right) \right]
      -2 \sin \tpl \sin  \left[ \tpl \left( -\frac p N + -\pn\right) \right].
\ee
This again gives a band of states which are cosine functions of the
flux, with period $L\Phi_0$. Differentiating $E_1$ wrt. 
$\alpha \equiv\tpl \left( -\frac p N - \pn\right) $, one finds the 
``amplitude'' of this band as
\be{eq:band}
E_{1,min} = -2 \left(1 + 2\cos \frac{4\pi}{L} + \cos \frac{2\pi}{L}  \right)
         \cos \alpha_{min}
      -2 \sin \tpl \sin  \alpha_{min}
\ee
\end{widetext}
with
\be{eq:band3}
\alpha_{min} = \tan^{-1} \left[ 
            \frac{\sin \tpl}{1 + 2\cos \frac{4\pi}{L} + \cos \frac{2\pi}{L}} \right].
\ee
Note that at {\it zero} flux, the lowest state of this band may have
an energy {\it larger} than $E_{1,min}$ -- the nearest minimum may
occur at a finite flux. This is because $p$ has to be an integer,
and $(LN/2\pi)\alpha_{min}$ is not generally an integer (except for
some special values of $L$).

The simplest example for four particles is $L=6$. In this case
$\alpha_{min}=\pi/3$, i.e. the energy of the state (\ref{eq:band1}) 
has a minimum at zero flux
with $p/N = -1$ and
$E_{1,min} = -\cos (\pi/3) - \sqrt{3} \sin(\pi/3) = -2$.

This method is easily generalized to construct higher bands. For
example, the second excitation band corresponds to, for four
particles,
$\{ I_j \} = -2, -1, 1, 2$
with energy
\be{eq:band4}
E_2   = -4 \left(\cos \frac{2\pi}{L} + \cos \frac{4\pi}{L}  \right)
         \cos \left[ \tpl \left( \frac p N + \pn\right) \right].
\ee

\begin{table}
\caption{Total weight of the most important configurations of 
the many-body states
of different vibrational states for the $t$-model ring with
$L=8$ and $N=4$. The configuration is shown as filled and empty
circles indicating whether or not there is a electron in 
the corresponding site.
The second column shows the number of such states (not including the 
degeneracy 
coming from the spin configuration). The last three colums show the total 
weights of these states for the vibrational ground state ($w_0$) and the
two vibrational bands. The total weights are the same for all rotational states
and different spin states belonging to the same vibrational mode.
}
\label{t1}
\begin{tabular}{crccc}
configuration & $n$ & $w_0$ & $w_1$ & $w_2$ \\
\hline
$\circ\bullet\circ\bullet\circ\bullet\circ\bullet$ & 8 & 0.1248 & 0.0000 & 0.0000 \\
$\circ\bullet\circ\bullet\bullet\circ\bullet\circ$ & 32 & 0.1824 & 0.2136 & 0.0624 \\
$\circ\bullet\bullet\circ\circ\bullet\bullet\circ$ & 16 & 0.0624 & 0.0000 & 0.1256 \\
\hline
\end{tabular}
\end{table}

The higher (inner) bands constructed in this way may
be interpreted as corresponding to vibrational excitations. 
As an example, we
have examined, for the $N=4$, $L=8$ solutions of the $t$-model,
which electron-hole configurations have the largest amplitude 
(in the many-body wavefunction) in
each of the three lowest bands. Table \ref{t1} shows the weights of
the most important basis states for the ground state and for the 
two lowest vibrational states. These are consistent with the 
classical motion of electrons in the first vibrational modes.

\subsection{Vibrational bands}
\label{s70.2}
In the limit $L \rightarrow \infty$, i.e. infinitely
many sites, the $t$-model is expected
to correspond to a system of non-interacting (spinless)
particles in the continuum (as they no longer ``see'' the delta function
interaction), in a similar way as the simple tight-binding model
approaches to the continuum model when the number of lattice sites
increases (see Sec. \ref{s60}). 
One way of illustrating that this is indeed the case, is to
examine the ratios
\be{eq:ratio}
R_n \equiv \frac{E_n - E_0}{E_1 - E_0},
\ee
where $E_i$ is the (minimum) energy of the $i$th excited band.
In the case of non-interacting, spinless particles the energy,
in units of $\hbar^2/(2 m_e R^2)$, of
an $N$-particle state at flux $\phi = \Phi/\Phi_0$ is given by
\be{eq:Efree1}
E = \sum_{i=1}^N  (m_i - \phi)^2
\ee
where $m_i$ are the single particle angular momenta and the total
angular momentum is $M = \sum m_i$. The minimum in energy of this
state occurs at $\phi = M/N$ and is given by 
$E_{min} = \sum_{i=1}^N m_i^2 -M^2/N$. 
As discussed in Sec. \ref{s10}, the ground state corresponds 
to filling consecutive angular
momentum states, e.g. $m_i = 0, 1, ..., N-1$. A set of $N/2$ excited
states is constructed by creating 1-hole excitations, the lowest
one being  $m_i = 0, 1, ..., N-2, N$. 
(Note that an overall shift of all the angular momenta does not 
change the minimum energy -- it just occurs at a different flux.)
Computing the ratios $R_n$ as defined in Eq.(\ref{eq:ratio})
one finds
\be{eq:freeratio}
R_n \equiv \frac{n(N-n)}{N-1}.
\ee
As we shall see, these one-hole excitations in a sense correspond to
fundamental phonon excitations. Moreover, it is easy to show that 2-hole
excitations lead to ratios (in the limit $N\rightarrow\infty$)
which are twice those in Eq.(\ref{eq:freeratio}), thus corresponding
to two-phonon excitations, etc.

Now, the same ratios can be recovered from the Bethe ansatz solution
of the $t$-model, as given by equations (\ref{eq:BA1}) and (\ref{eq:BA2}): 
Constructing the set of excitations which correspond
to one hole in the quantum numbers $I_j$, minimizing the energy as
described in the previous section, and taking the limit
$L\rightarrow\infty$,
one again gets Eq.(\ref{eq:freeratio}). This shows that the energy
bands of the Hubbard model reduce to those of non-interacting
particles in the
limit of infinite repulsion and infinite number of sites.
(Of course the correspondence between the $t$-model and free particles
can also be seen using a similar argument as in section \ref{s60}: 
For the maximum spin state, $-p = \sum J_{\alpha} = 0$ in
Eq.(\ref{eq:M}) so that the integers $I_j$ correspond 
just to the single particle 
angular momenta.)

\subsection{$1/d_{ij}^2$ interactions:}
\label{s70.3}

There is another type of interaction in one dimension which leads
to the same set of excitation bands as above, namely $V_{ij}=1/d_{ij}^2$, 
where $d_{ij}$ is the distance between particles $i$ and $j$.
Models in 1D with this type of interaction are known
as the Calogero- (on a line) or Sutherland model (on a circle)
\cite{calogero1969,sutherland1971} and 
have been studied extensively over the past three decades. 
What makes the Calogero (or Sutherland) model special, is that
it mimics as closely as possible a system of free particles.
Consider the $N$-particle Sutherland Hamiltonian,
\be{eq:SM}
H_S = -\sum_{i=1}^N \frac{\partial^2}{\partial x_i^2} 
    + 2\lambda(\lambda-1)\sum_{i<j}
    \frac{1}{\left(2\sin\left( \frac{x_i-x_j}{2} \right)\right)^2},
\ee
where we have set the radius of the circle to $1$. The complete
excitation spectrum of this model can be found exactly in terms
of the asymptotic Bethe ansatz (see e.g. \cite{shastry1994} and
references therein) and is remarkably simple.
The total energy and angular momentum of a given state
can be written as $E = \sum_j p_j^2$ and $P=\sum_j p_j$, respectively,
where the quasi-momenta $p_n$ are related to the free angular
momenta $m_j$ by
\be{eq:BA}
p_j = m_j + \gamma\left( j - \frac{N+1}{2} \right),
\ee
where $\gamma = (\lambda - 4 )/4 $.
In other words, the energy and total angular momentum take the
same form as in the non-interacing case, with the interactions
absorbed in the shift of the quasi-momenta. For each non-interacting
many-body state characterized by a set of fermionic quantum
numbers $\{ m_j \}$ there is a
corresponding state with quantum numbers $\{ p_j \}$. Note that
$P=\sum_j p_j = \sum_j m_j = M$.
Now consider the Sutherland model in the presence of an
Aharonov-Bohm flux piercing the ring. The energy of a given state
can then be written as 
\be{eq:ABSM}
E = \sum_j (p_j - \phi)^2
\ee
and, as in the non-interacting case, is minimized for $\phi=M/N$,
\be{eq:ABSMmin}
E_{min} = \sum_j p_j^2 - M^2/N.
\ee
We want to compute the shift in energy of a many-body state as
compared to the non-interacting one, which then will be used to
compute the ratios $R_n$. Inserting the solution (\ref{eq:BA}) into
Eq.(\ref{eq:ABSM}) one finds, up to an overall constant,
\be{eq:Eshift}
\Delta E(\gamma) &=& E_{min}(\gamma) - E_{min}(\gamma = 0) \\
      &=& \gamma \left( \sum_{j=1}^N j m_j - (N+1) M \right).
\ee 
Again constructing the set of one-hole excitations (in $m_j$)
as before, one easily finds the {\em difference} between the shift
of the $n$th excitation and the shift in the ground state,
\be{eq:relshift}
\Delta E^{(n)}(\gamma) - \Delta E^{(0)}(\gamma)
      = \gamma \left( n(N-n) \right).
\ee 
From this it immediately follows that the ratios $R_n$ for the
Sutherland model are the same as
in the non-interacting case, see Eq.(\ref{eq:freeratio}).
In addition, recall section \ref{s30} where we performed a calculation
of the normal modes for a set of {\em classical} particles
on a ring with $1/d^2$ repulsion. In a semiclassical picture, the 
corresponding frequencies $\omega_j$  were then interpreted as the 
eigenfrequencies of the many-body problem, with energies $\hbar\omega_j$.
Using these to compute the ratios $R_n$, one again obtains the same 
expression as above. This illustrates that
the ``single-particle excitations'', both in the non-interacting
case and in the Sutherland model, can indeed be interpreted as
corresponding to vibrational modes. (Cfr. the discussion of the
free particle case in Sec. \ref{freevibsec}.)

\subsection{Finite $U$}
\label{s80}

The Hubbard model with finite $U$ can still be solved
with the Bethe ansatz but now it requires numerical
solution of the set of nonlinear equations (\ref{eq:bethe1}). 
In the case of
small number of electrons and sites a direct 
numerical diagononalization
of the Hubbard Hamiltonian, Eq. (\ref{eq:hub}), is in fact easier.
The results shown in this section have been computed
with the direct numerical diagonalization. As an
example case we use again the 4 electron ring.

\begin{figure}
\includegraphics[angle=-90]{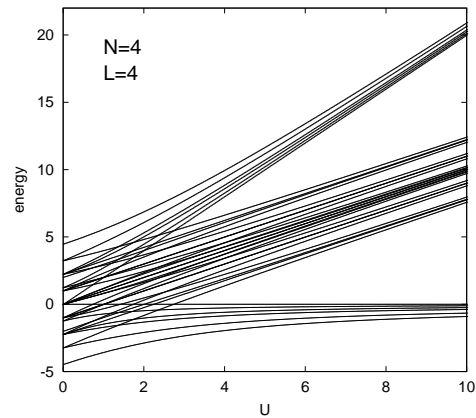}
\caption{Energy levels as a function of $U$ for 
a four electron Hubbard ring with four sites.      
}
\label{f8}
\end{figure}

Figure \ref{f8} shows the energy spectrum for
four electrons in four sites as a function of 
on-site interaction $U$. The results start from the 
noninteracting case (with spin). Increasing $U$
separates the spectrum into different groups. The
lowest group corresponds to states where all the 
electrons are mainly at different lattice sites while 
the two higher bands correspond to states where one 
or two electrons are at the same site, respectively. 
This can be seen 
by looking at the structure of the many-body states or 
simply by noticing that the energy of the higher
groups of levels increase as $E\approx U$ and $E\approx 2U$.

The high-$U$ limit of the lowest group of states can be 
explained with the so-called $tJ$-model.
In the limit of large $U$ the half-filled 
Hubbard model can be approximated as (see \cite{vollhardt1994}
and references therein)
\begin{equation}
H_{tJ}=H_t+\frac{J}{2}\sum_{i\neq j}^N 
\left( {\bf S}_i\cdot{\bf S}_j-\frac{1}{4}\right),
\label{fu1}
\end{equation}
where $H_t$ denotes the Hamiltonian of the 
infinite $U$ limit ($t$ model), $J=4t^2/U$ and 
${\bf S}_i$ is a spin operator ($S=1/2$). 
In the case of a half-filled band the $t$-model gives 
only one state (with zero energy). In the 
large $U$ limit all the low energy
states can then be described with an
antiferromagnetic Heisenberg Hamiltonian
which separates the different spin-states.
The case with four electrons is especially easy to solve
(see exercise 30.3. of Ref. \cite{ashcroft1976}),
leading to energy levels shown in Table \ref{t2}.
The table also shows the orbital angular momentum determined 
from the symmetry properties of the Heisenberg state.

\begin{table}
\caption{Energy states of a Heisenberg ring with four
electrons, sorted according to the spin $S$ and total
angular momentum $M$ quantum numbers. $M$ is determined 
with the help of the symmetry of the state.}
\label{t2}
\begin{tabular}{ccrr}
$S$ & $S_z$ & $M$ & $E$ \\
\hline
0 & 0 & 2 & -2J \\
0 & 0 & 0 & -J \\
1 & -1,0,1 & 0 & 0 \\
1 & -1,0,1 & 1 & 0 \\
1 & -1,0,1 & 3 & 0 \\
2 & -2,-1,0,1,2 & 2 &J \\
\hline
\end{tabular}
\end{table}

\begin{figure}
\includegraphics[angle=-90]{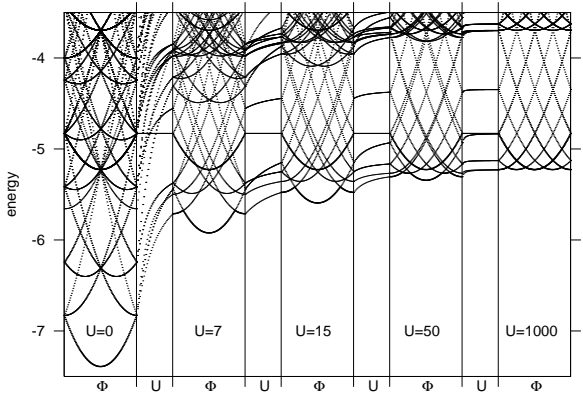}
\caption{Flux and $U$ dependence of the many-body states 
of the Hubbard model with $N=4$ and $L=8$. For each 
fixed $U$ the flux goes from 0 to $\Phi_0$; in between,
$\Phi=0$ and $U$ increases linearly to the next fixed value.
The lowest overall energy and the lowest energy state 
corresponding to the maximum spin are shown as thick lines.
Note that the maximum spin state is independent of $U$ and that
the periodicity of the yrast state changes from $\Phi_0$ to
$\Phi_0/4$ when $U$ increases from 0 to $\infty$.
}
\label{f9}
\end{figure}

In the case of a large $U$ the magnetic flux does
not have any effect if $N=L$ due to the fact that the 
electron motion is strongly hindered. The situation changes
if empty sites are added. Figure \ref{f9} shows the 
development of the low energy states as a function of the 
magnetic field and $U$ in the case of four electrons in 
eight sites. When $U$ is reduced from infinity, the 
different spin-states separate, causing the periodicity 
of the yrast line to change from $\Phi_0/N$ at $U=\infty$
to $\Phi_0$ at $U=0$. This change of the periodicity is
addressed in more detail in Sec. \ref{s110}.

The effect of finite, but large $U$, is to split the degeneracy
of the different spin-states of the $t$-model. If there are no
empty sites, the Hamiltonian approaches that of the $tJ$-model,
Eq. (\ref{fu1}).  We have solved the spectrum of four electrons
as a function of the flux,
increasing the number of sites from 4 to 12.
The results show that when the number of empty sites increases,
the spectra are still in fair agreement with those of the $tJ$-Hamiltonian,
but the effective coupling between the spins, $J$, is not any
more $4t^2/U$, but decreases rapidly when the number of empty sites
increases.

\begin{figure}
\includegraphics[angle=-90]{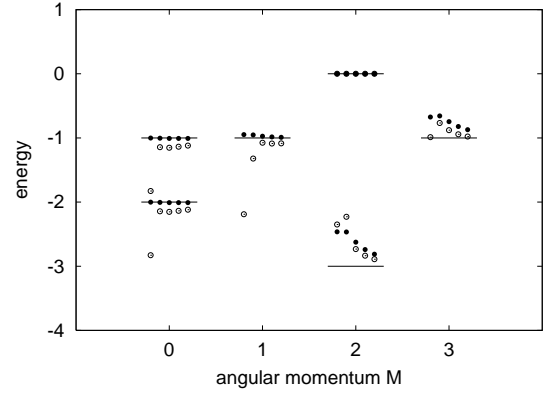}
\caption{The Heisenberg model energy spectrum 
for four electrons determined from the
Hubbard model with empty sites. The black dots 
are the results for four electrons with $U=100$ and 
different number of sites: For each state the dots
from left to right correspond to  $L=5$, 6,
8, 10, and 12, respectively. For comparison, the solid lines
denote the half-filling case (no empty sites), 
i.e. $L=4$. In each case the energy
difference of the two $M=0$ states is adjusted to be 1.
Open circles show the same for $U=10$.
}
\label{f10}
\end{figure}

Figure \ref{f10} shows the difference of the energy spectra derived 
from the exact Hubbard spectrum and from the $t$-model, for 
different values of empty sites. The results are scaled so that
the energy difference between the two spin states corresponding
to $M=0$ are the same. We can see that when the number of empty sites
increases, the difference spectrum approaches that of the 
Heisenberg Hamiltonian. These results suggest that the 
separation of the Hubbard Hamiltonian (for large $U$) into the
$t$-model and Heisenberg Hamiltonians is accurate in the 
two limits, $N/L\rightarrow 1$ and $N/L\rightarrow \infty$.
Moreover, the agreement with {\it all} numbers
of empty sites is surprisingly good, even if $U$ is as small as 10.

\begin{figure}
\includegraphics[angle=-90]{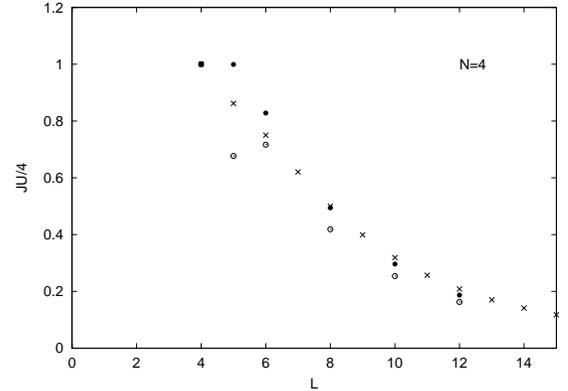}
\caption{The dependence of the effective Heisenberg coupling $J$
on the number of sites ($L$). The crosses show the asymptotic
large $U$ limit of Eq. (\ref{jeff}), the black dots (open circles)
the result determined from the exact solution for $U=100$ 
($U=10$).
}
\label{f11}
\end{figure}

Yu and Fowler\cite{yu1992} have used the Bethe ansatz to study the 
large $U$ limit for any number of empty sites and shown that the
effective $J_{\rm eff}$ is related to the quantum numbers $k_j$ as
\begin{equation} 
J_{\rm eff}=\frac{4}{LU}\sum_{j=1}^N \sin^2 k_j.
\label{jeff}
\end{equation}
Figure \ref{f11} shows the effective coupling constant
determined by this equation as a 
function of the number of sites for a four electron ring,
compared to those obtained from the direct diagonalization
of the Hubbard Hamiltonian for $U=10$ and $U=100$.
The agreement is fairly good for any number of empty sites 
even for the relatively small $U=10$.
The decrease of $J_{\rm eff}$ as a function of $L$ can be 
explained with electron 'localization': When the number of empty
sites increases, the localized electrons move further apart
reducing the exchange interaction.
We find, once again, that in the 1D system the electrons with 
repulsive (even $\delta$-function) interactions
behave as localized particles.

\section{Quasi-1D-continuum rings: Exact CI method and effective Hamiltonian}
\label{s90}
In this section we review electronic structure calculations for 
quasi-1D rings, published in 
Refs. \cite{koskinen2001,koskinenp2002}; they
were a natural continuation of earlier calculations done for 
electrons in harmonic two-dimensional quantum dots\cite{reimann2002}.
Usually the quantum ring is described 
with a displaced harmonic confinement
(although several other models have been used\cite{niemela1996,tan1999})
\begin{equation}
V(r)=\frac{1}{2}m_e\omega_0^2(r-r_0)^2,
\label{q1dr}
\end{equation}
where $r_0$ is the radius of the ring and $\omega_0$ the perpendicular
frequency of the 1D wire. Note that we still assume the ring
to be strictly two-dimensional, i.e. it is infinitely thin
in the direction perpendicular
to the plane of the ring. The parameters describing the ring,
$r_0$ and $\omega_0$ can be related to the density parameter 
$r_s=1/(2 n_0)$ of the 1D system ($n_0$ is the 1D density) and to
a parameter describing the degree of one-dimensionality of the wire.
For the latter, Reimann {\it et al.}\cite{reimann1999,koskinen2001} used 
a parameter $C_F$ defined with the relation 
$\hbar\omega_0=C_F\hbar^2\pi^2/(32mr_s^2)$. The physical meaning of
$C_F$ is that it is the ratio of the first radial excitation in the ring 
to the Fermi energy (approximated by that of an ideal 1D Fermi gas with
the same density) of the 1D electron gas.

There are several approaches to solve the many-body problem 
of interacting electrons in the above potential. In studying the 
spectral properties the most useful method is the brute force
diagonalization of the many-body Hamiltonian in a proper basis.
This Configuration-Interaction (CI) 
method gives the whole many-body energy spectrum as well 
as the corresponding many-body states. Naturally, the solution 
can be numerically accurate only for small numbers of electrons,
typically less than 10. The matrix dimension can be
reduced by fixing the orbital angular momentum to the desired value.
For example, Koskinen {\it et al.}\cite{koskinen2001} first expanded the
solutions of the single particle part of the Hamiltonian in a harmonic
oscillator basis and then used these functions as a
single particle basis for the Fock space in doing the CI calculations.
According to their eigenenergies, up to 50 single-particle
states were selected to span the Fock space and the number of
the many-body Fock-states was restricted to about $10^5$ using another
energy cutoff (for a given total $M$).
For an even number of electrons all spin-states can be
obtained with fixing $N_\uparrow=N_\downarrow=N/2$. The total spin
of each state can afterwards be determined by calculating the expectation
value of the $\hat S^2$ operator.

\begin{figure}
\includegraphics{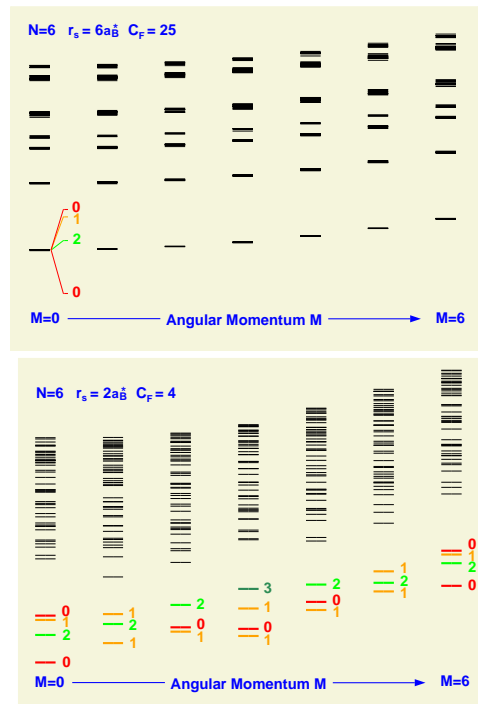}
\caption{Energy spectra for two quasi-one-dimensional
continuum rings with six electrons (in zero magnetic field).
The upper panel is for a narrow ring and it shows several
vibrational bands. The lower panel is for a wider ring 
which shows stronger separation of energy levels corresponding
to different spin states (shown as numbers next to the energy levels). 
Note that also the narrow ring has the same spin-ordering of the 
nearly degenerate state as expanded for the lowest $M=0$ state.
}
\label{f12}
\end{figure}

Figure \ref{f12} shows the calculated energy spectra obtained
with such a calculation for two different
rings with six electrons in each. It is instructive to introduce
a simple model Hamiltonian\cite{koskinen2001}
\begin{equation}
H_{eff} = \frac{\hbar^2}{2I} {\bf M}^2 
        + \sum_{\nu} \hbar \omega_{\nu} n_{\nu}
	+ J\sum_{\langle i,j\rangle} {\bf S}_i\cdot{\bf S}_j
\label{eq:ModHam}
\end{equation}
where $\omega_{\nu}$ are the frequencies and the integers 
$n_{\nu}$ the number of excitation quanta of the various vibrational
normal modes, and $I$ is the moment of inertia of the `molecule'. 
This Hamiltonian is thus simply a combination of rigid rotation
of the whole system, internal vibration, and a Heisenberg term to capture 
the spin dynamics,
and one may examine how well it describes the exact results.
To this end, note the following interesting features in Fig. \ref{f12}: 
The narrower ring shows clear rotation-vibration 
bands, very similar to those obtained for electrons in a continuum
ring with $\delta$-function interaction, cfr. Fig. \ref{f5}. 
The only difference is that
now the ratios of the vibrational states correspond to those 
determined for classical electrons interacting with $1/r$ interaction.

Each spectral line consists of several nearly degenerate spin-states.
The inset shows as an example the detailed structure of the $M=0$
state. The spin structure coincides with that determined from the 
$tJ$-model, i.e. it can be determined by solving the antiferromagnetic
Heisenberg model for a ring of six electrons. The ratios of the 
energies of different spin-states are quantitatively the same as
in the Heisenberg model. 

The lower panel of Figure \ref{f12} shows that this is true also
for a wider ring. In this case only the vibrational ground state 
is clearly separated from the rest of the spectrum. However,
the internal structure of this yrast band is still very close
to that of the Heisenberg model: Qualitatively the agreement is
exact, i.e. each angular momentum has the right spin states in the 
right order. Only the energy ratios are not any more exactly 
the same as in the Heisenberg model. Koskinen et al.\cite{koskinenp2002}
have studied in more detail how well the model Hamiltonian
(\ref{eq:ModHam}) 
describes the exact many-body results for a ring of six electrons.

\begin{figure}
\includegraphics{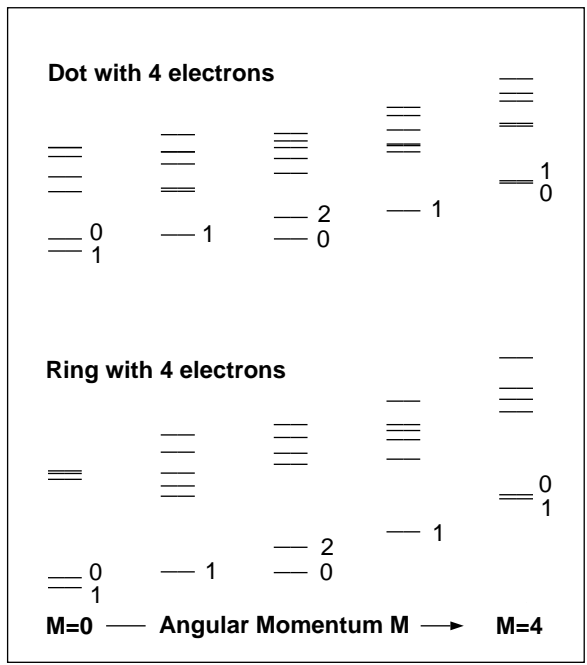}
\caption{Low energy spectra of a four electron quantum dot (upper panel) and
a four electron quantum ring (lower panel) as a function of the angular 
momentum.
The numbers next to the energy levels give the total spin.
CI calculation from \cite{manninen2001}.
}
\label{f13}
\end{figure}

Exact CI calculations for a ring of 4\cite{manninen2001}, 
and 5 and 7\cite{koskinen2001}
electrons show similar agreement. 
Figure \ref{f13} shows the energy spectra for a four electron ring with a
comparison to a four electron dot. With such a small number of electrons
even the dot shows nearly the same yrast spectrum as the ring. 
This indicates that also in the dot the electrons will localize 
in a square Wigner molecule (see also \cite{reimann2000}).
Similar localization of electrons and their rotational and vibrational 
spectra are observed also in two-electron quantum 
dots\cite{yannouleas2000} 
(where the energy spectrum can be solved exactly\cite{dineykhan1997}).
When the number of electrons in a quantum {\it dot} is 6 or 
larger the classical configuration
of electrons in the Wigner molecule\cite{bolton1993} 
is not any more a single ring and the 
spectral properties become more complicated. Nevertheless, even there
the polarized case shows rotational bands consistent with the localization of
electrons\cite{manninen2001b}.
We should point out that the idea of describing few-electron
systems in terms of rigid rotation and internal vibrations,
is not new, but was first applied to quantum dots by 
Maksym\cite{maksym1996}.

\section{Exact diagonalization: Finite magnetic field}
\label{s100}
\subsection*{Flux inside the ring}

Let us first consider a magnetic field concentrating
in the center of the ring, as described with the 
vector potential (\ref{noa1}). Since the $r$-component
of the vector potential is zero and the $\varphi$-component
is proportional to $1/r$, the only effect
of the flux is to change the
angular momentum term of the single particle
Schr\"odinger equation as
\begin{equation}
\frac{\hbar^2m^2}{2m_er^2}\rightarrow
\frac{\hbar^2(m-\Phi/\Phi_0)^2}{2m_er^2}.
\label{eq100_1}
\end{equation}
It is easy to see that for integer $\Phi/\Phi_0$ the energy
levels and single particle states are equivalent to those
without the flux, they are just shifted to another
angular momentum value. Since each Slater
determinant of the many-body state has a good angular
momentum quantum number, the same is true for the many-body state:
The angular
momentum $M$ is shifted to that of $M-N\nu$ when a integer
number $\nu=\Phi/\Phi_0$ of flux quanta is penetrating the ring
(as in Eq. (\ref{ni9}) for a strictly 1D ring).
This means that we get the same picture as discussed before:
The effect of the increasing flux is just to tilt the spectrum so
that successively higher and higher angular momentum values 
become the ground state. 
The ground state energy will be periodic with flux:
$E_M(\Phi)=E_{M+N\nu}(\Phi+\nu\Phi_0)$.
Note that this is true even if the 
$M$-dependence of the many-body spectrum is not exactly
proportional to $M^2$; the only requirement is that the 
flux is restricted in the central region of the ring and
the magnetic field does not overlap with the single particle states.
In this case we can write the model Hamiltonian describing the 
yrast spectrum of the exact CI calculation as
\begin{equation}
H=J\sum_{\langle i,j\rangle } {\bf S}_i\cdot{\bf S}_j
+\frac{\hbar^2}{2Nm_eR^2}\left(M-\frac{N\Phi}{\Phi_0}\right)^2.
\label{eq100_2}
\end{equation}
The observed periodicity of the ground state energy, or persistent current,
as a function of flux is determined by the variation of the  
yrast-line as a function of $M$, for example as shown in Fig.
\ref{f12}. If the situation is like in the upper panel of 
Fig. \ref{f12}, i.e. when the ring is very narrow, the energy
increases accurately as $M^2$. In this case Eq.(\ref{eq100_2})
gives the periodicity $\Phi_0/N$. On the other hand in the case
of the lower panel of Fig. \ref{f12} the minimum energy 
jumps from $M=0$ to $M=3$ and then to $M=6$, when the flux
is increased. This means a periodicity of $\Phi_0/2$.
If the ring is made even wider, eventually the minimum at $M=3$
will not be reached and the periodicity changes to $\Phi_0$.

Finally, we should notice that if the electron gas were
polarized (spinless electrons), the periodicity would always be
$\Phi_0$. This can be seen from
Fig. \ref{f12} which shows that the maximum spin state 
occurs in the lowest vibrational band only at the angular 
momentum $M=3$ (more generally, at $M=N/2$ for even number of
particles and at $M=0$ for odd number of particles).

\subsection*{Homogeneous magnetic field, no Zeeman splitting}

\begin{figure}
\includegraphics{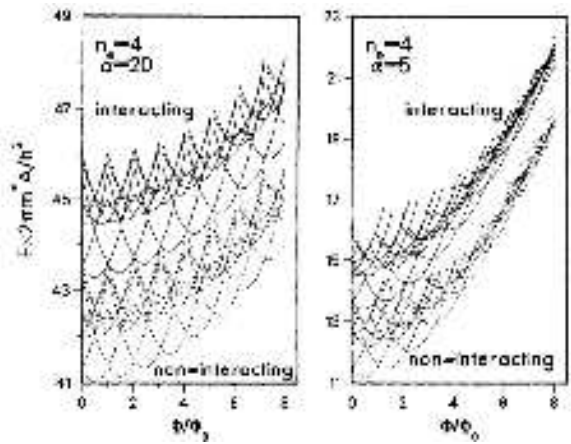}
\caption{Energy levels for four noninteracting (dashed lines)
and interacting (solid lines) spinless electrons in a 
ring as a function of the magnetic flux. The magnetic field is homogeneous.
From Ref. \cite{chakraborty1994}.
}
\label{f14}
\end{figure}

Experimentally it would be easier to measure the quantum rings
in the presence of a homogeneous magnetic field. 
This case was first treated by first principle
calculation methods in the pioneering papers
by Chakraborty and Pietil\"ainen\cite{chakraborty1994}
and by Niemel\"a et al\cite{niemela1996}, and the present
section is mainly based on these papers.
In this case the
vector potential can be expressed, for example, 
in terms of a symmetric gauge 
${\bf A}=\frac{1}{2}(-By,Bx,0)$.
This vector potential effectively adds an additional
harmonic confinement centered at the origin. 
The $r$-dependent single particle potential changes as
\begin{equation}
\frac{1}{2}m_e\omega_0^2(r-r_0)^2\rightarrow
\frac{1}{2}m_e\omega_B^2(r-r_B)^2+{\rm constant}
\label{eq100_3}
\end{equation}
where $\omega_B^2=\omega_0^2+e^2B^2/(4m_e^2)$ and 
$r_B=r_0\omega_0^2/\omega_B^2$. 
The effect of the field is then just to change the
parameters of the confining ring. This will change the 
energy differences between the single particle states,
but as long as the ring is narrow enough to have only
one radial mode, it can not change their order.
Consequently, in narrow rings the effect of the field
on the many-body state is small and only quantitative.
Nevertheless, the change of the potential shape and the 
constant term means that the lowest single particle
state increases with the flux\cite{chakraborty1994}, as shown in Fig. 
\ref{f14} for spinless electrons.
The figure also indicates that the effect of the 
electron-electron interactions in the spinless case
is mainly to shift the spectrum upwards by a constant.

\begin{figure}
\includegraphics{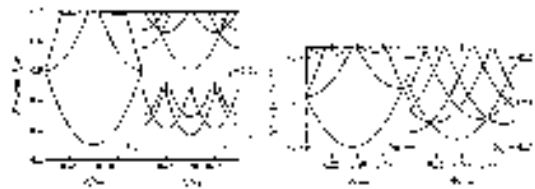}
\caption{Energy levels as a function of the magnetic flux in a ring of 
four electrons. The left-hand side shows the results for delta-function
interaction in a strictly 1D ring, (a) for noninteracting, 
(b) and (c) for interacting so that
in (c) the interaction is twice as strong as in (b). 
The right hand panel shows the result for a 
quasi-1D ring with $1/r$-interaction,
(c) noninteracting and (d) interacting electrons. Note the
similarity of the results
in (b) on the left-hand side and (d) on the right hand side.
From Ref. \cite{niemela1996}
}
\label{f15}
\end{figure}

Niemel\"a {\it et al.}\cite{niemela1996} 
have performed an extensive study of quasi-1D-rings
with 2 to 4 interacting electrons. They used a homogeneous magnetic
field and neglected the Zeeman splitting. The results for two and 
three interacting electrons show periodicities $\Phi_0/2$ and
$\Phi_0/3$, respectively, i.e. consistent with the $\Phi_0/N$
periodicity for a narrow ring. In the case of four electrons 
Niemel\"a {\it et al.} studied in addition to the $1/r$ Coulomb
interaction also a $\delta$-function interaction ($V_0\delta(r)$)
in the case of an infinitely narrow ring. 
It is then interesting to compare the 
results of these two models, as shown in Figure \ref{f15}.
For the $\delta$-function interaction the results are
expected to be the same as for the Hubbard model with an infinite number
of lattice sites. Indeed, the results show the change of periodicity
from $\Phi_0$ first to $\Phi_0/2$ and then to $\Phi_0/4$, when the 
strength of the $\delta$-function interaction ($V_0$) is increased.
The results of  Fig. \ref{f15} compare well with the results of 
the Hubbard model with only 8 sites, i.e. those shown in Fig. \ref{f9} 
for $U=0$, $U=15$ and $U=50$.

The right hand side of Fig. \ref{f15} shows the energy levels for a
quasi-1D-ring
with $1/r$-interaction. The spectra are essentially the same, the
only difference being a slight upward shift of the energy when the flux
changes from 0 to 1. This is due to the harmonic repulsion of the 
flux-dependent effective potential, Eq. (\ref{eq100_3}), caused by the 
homogeneous magnetic field. Note the similarities of the spectra shown in
Figs. \ref{f15} b and c.

\section{Periodicity of persistent current}
\label{s110}
\subsection{Strictly 1D rings: Spectrum of rigid rotation}

The many-electron excitation spectrum for electrons interacting
with the infinitely strong $\delta$-function interaction was
studied using the Hubbard model in Sec.\ref{s70}, see Fig. \ref{f7}.
As mentioned, this spectrum can be constructed from the Bethe ansatz.
In the low-energy part of the spectrum (close to the yrast line)
the levels consist only of the compact states of the 
quantum numbers $I_j$, i.e. there is no vibrational
energy. It is instructive to
use the continuum limit ($L/N\rightarrow \infty)$)
of the Bethe ansatz solution for the $t$-model
to compare these (yrast) energy levels as a function of the flux $\Phi$
to those of {\it single} electron states in a ring of radius $R$.
One finds for a many-electron state
\begin{equation}
E(M,\Phi)=\frac{\hbar^2}{2Nm_eR^2}\left(M-\frac{\Phi}{h/Ne}\right)^2
\quad (+ constant)
\label{e1101}
\end{equation}
and for a single electron state
\begin{equation}
\epsilon(m,\Phi)=\frac{\hbar^2}{2m_eR^2}\left(m-\frac{\Phi}{h/e}\right)^2
\label{e1102}
\end{equation}
where we have now written the flux quantum as $\Phi_0=h/e$.
We notice that the many-electron states are identical to the 
single electron levels with the electron mass and charge,
$m_e$ and $e$, replaced by 
the total mass and total charge of all the electrons, $Nm_e$ and $Ne$.
Indeed the strongly correlated electron system with infinitely
strong delta function interaction behaves as a rigidly rotating
single particle. 
We should note that in the spinless case the number $p$ of 
Eq.(\ref{eq:BA2}) is always zero and 
consequently one recovers the noninteracting case (for spinless electrons the 
$\delta$-function interaction does not have any meaning due to the 
Pauli exclusion
principle). The rigid rotation of the electron system then leads always to 
the $\Phi_0$ periodicity as discussed already in Sec.\ref{s10}

We have learned in Secs. \ref{s90} and \ref{s100} 
on the basis of numerical solutions
for narrow quasi-1D rings, that electrons with normal $1/r$ interactions
also produce similar rotational spectrum. Moreover, the solution of the 
Calogero-Sutherland model shows rigorously that a similar 
spectrum is observed for $1/r^2$-interaction. Both these
interactions have the property that they are infinitely strong
at the contact, preventing the electron to
pass each other. It seems obvious that any repulsive
interaction which is infinitely strong (in such a way that 
electrons with opposite spin are not allowed at the same 
point) produces for the strictly 1D ring the same spectrum
of rigid rotation. 

The yrast spectrum is qualitatively similar for all electron
numbers. In particular, in experimentally determined spectra,
the number of electrons in the ring can 
be seen as a qualitative change of the spectrum
only by observing the vibrational bands.
(The yrast energy alone determines the number of electrons 
in a narrow ring only if the flux is quantitatively determined.)
A way of estimating the number of particles is to count the
number of (purely rotational) states below the onset of the first
vibrational band, as can be seen from the following argument:
By considering noninteracting spinless
electrons (which have exactly the same energy levels, though not all
of them, as spinful electrons) we know from Eq.(\ref{ni5})
that the first vibrational level is the 
first noncompact state, having an excitation energy of about
$N\hbar^2/2m_eR^2$. This equals the $E(M=N,\Phi=0)$ energy
of the yrast band, Eq.(\ref{e1101}). For $N$ particles there are thus about
$N$ purely rotational states below the first vibrational band.
Figure \ref{f7} demonstrates that this is true for $N=4$.

\subsection{Periodicity change in quasi-1D rings}

In this section we will concentrate on the lowest energy 
state (the yrast state) and study in more detail its periodicity 
in quasi-1D rings, where the electrons are allowed to pass each other.
(The periodicity of the persistent current at zero temperature
is the same as that of the ground state energy).
As discussed in Sec. \ref{s80}, the (strictly 1D) 
Hubbard model suggests that
(i) for spinless electrons the periodicity is always the flux
quantum $\Phi_0$ and (ii) for electrons with spin the periodicity
changes from $\Phi_0$ first to $\Phi_0/2$ and then to $\Phi_0/N$
when the interaction $U$ increases from zero to infinity,
as illustrated in Fig. \ref{f9}. This change of periodicity in the 
Hubbard model was first studied by Yu and Fowler\cite{yu1992}
and Kusmartsev {\it et al.}\cite{kusmartsev1994}.

\begin{figure}
\includegraphics[angle=-90]{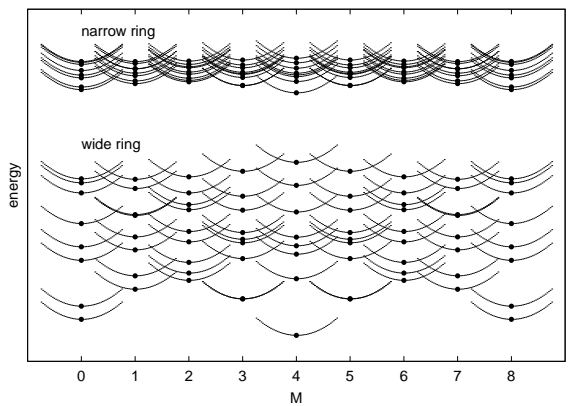}
\caption{Schematic illustration of the $\Phi_0/N$ and $\Phi_0/2$ periodicities
of a ring with eight electrons. The black points show the energy spectra with
two different values of $J$. The parabolas drawn at each point indicate 
the change 
of the rotational energy level as a function of the magnetic flux. The lowest
envelope of the overlapping parabolas gives the
ground state energy as a function of the flux.
}
\label{f16}
\end{figure}

The change in periodicity can be traced back to the notion
that the Hubbard model with empty sites can be quite accurately
described by the $tJ_{\rm eff}$-model with an effective exchange coupling 
$J_{\rm eff}$ which depends not only on $U$ but also on the number of 
empty sites,
as demonstrated in Figs. \ref{f10} and \ref{f11}. When $J_{\rm eff}$ is small, 
the $\Phi_0/N$ periodicity is observed, while for large $J_{\rm eff}$
the $\Phi_0$ periodicity is found (for even numbers of electrons).
The periodicity of $\Phi_0/2$ results from the fact that the 
solution of the Heisenberg Hamiltonian has two close lying states
corresponding to angular momenta $M=N/2$ and $M=0$, while the states
corresponding to other angular momenta are clearly higher in energy. 
The situation is demonstrated in Fig. \ref{f16} where we show the 
energy spectrum for the Heisenberg model for an eight electron ring.
The energy spectrum for the total Hamiltonian as a function of the 
flux can be estimated simply by drawing a parabola at each energy level
as shown in the figure. Now depending on the 
energy splitting of the Heisenberg model, i.e. on  $J_{\rm eff}$, 
the resulting lowest energy state as a function of flux can have
periodicities, $\Phi_0/N$, $\Phi_0/2$, or $\Phi_0$.

Deo {\it et al.}\cite{deo2003} have studied the periodicity change using the 
model Hamiltonian Eq.(\ref{eq100_2})
which is known to give good agreement with exact diagonalization results
of the quasi-1D continuum Hamiltonian. 
This model Hamiltonian has the same flux dependence as Eq.(\ref{e1101}),
which was derived from the Hubbard model.
It is thus natural that the same
periodicity change is observed. 
In both cases the flux only affects the rotational state of 
the system by changing its energy as
$M^2 \rightarrow (M-N\Phi/\Phi_0)^2$,
but does not change the {\it internal structure} of the many-body
state in question, i.e. all interparticle correlations
remain the same. 

\begin{figure}
\includegraphics[angle=-90]{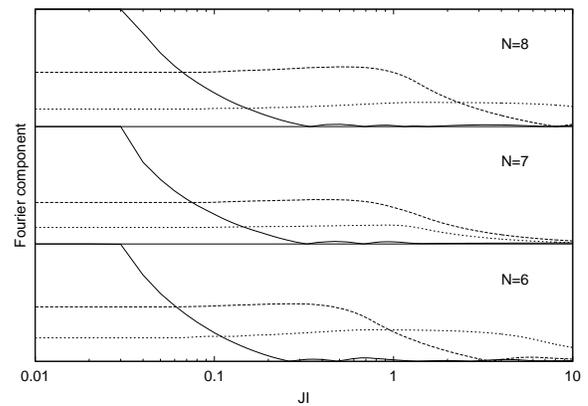}
\caption{Three largest Fourier components of the periodic ground state energy. 
$JI$ is the product of the effective Heisenberg 
coupling and the moment of inertia.
The dotted, dashed and solid lines are the Fourier components corresponding to
the periodicities $\Phi_0$,  $\Phi_0/2$ and $\Phi_0/N$, respectively.
}
\label{f17}
\end{figure}

We can look at the periodicity change in more detail
by making a Fourier analysis of the lowest energy state.
We have done this using the model Hamiltonian (\ref{eq100_2}).
A parameter determining the relative weights of the 
energy states of the two separable parts of the Hamiltonian is 
then $JI$, i.e. the product of the moment of inertia ($I\approx Nm_eR^2$)
and the Heisenberg coupling parameter
(in the Hubbard model the corresponding parameter would be $J_{\rm eff}$). 
Figure \ref{f17}
shows the three most important Fourier components as a function
of $JI$ for rings with 6, 7 and 8 electrons. For 6 and 8 electrons
we see clearly the period changes. In the case of an odd number 
of electrons, the solution of the Heisenberg model is qualitatively
different: There are two degenerate minima corresponding to 
different angular momenta. Consequently, the period $\Phi_0/2$ stays
always more important than the period $\Phi_0$\cite{deo2003}. 

We have learned that the periodicity change as a function of the 
effective width of the quasi-one-dimensional continuum ring 
is similar to that in a strictly 1D Hubbard ring as a function 
of $U$, suggesting that the finite $U$ in the Hubbard model
mimics the finite width of a continuum ring. 
This similarity can be understood as follows.
With infinite $U$ the electrons are forbidden to pass each other.
This situation is similar to an infinitely narrow continuum ring 
with $1/r$ interaction between the electrons. When $U$ gets smaller,
the electrons (with opposite spin) can hop over each other, the better the
smaller $U$ is. Naturally, in the continuum ring the electrons 
are allowed to pass each other 
if the ring has a finite width.
Decreasing $U$ in the Hubbard model thus corresponds
to making the continuum ring wider.

Note that in order to see the periodicity in the Hubbard model,
one has to consider systems where the number of electrons $N$ is 
smaller than the number of lattice sites $L$, i.e. empty sites are 
needed for 'free rotation' of the ring.
The resulting periodicity depends both on the number of empty sites
and on the on-site energy $U$.

\section{Other many-body approaches for quasi-1D continuum rings}
\label{s120}

\subsection*{Quantum monte carlo}

A class of powerful tools to study the low energy states of 
many-body quantum systems are based on the Monte Carlo 
method\cite{ceperley1996}.
These methods have been extensively applied also for
quantum dots\cite{bolton1994,egger1999,harju1999,shumway2000,colletti2002} 
and quantum rings\cite{borrmann2001,pederiva2002} with a few electrons.
Monte Carlo approaches 
are most suitable for studying either the ground state properties
(variational Monte Carlo and diffusion Monte Carlo) or average finite 
temperature properties (path integral Monte Carlo), 
and have thus not been able to produce the detailed spectral properties 
with the accuracy of the CI method. It also seems that even finding the 
correct ground state is not straightforward by using Quantum Monte 
Carlo\cite{qmproblem}

Pederiva {\it et al.}\cite{pederiva2002} have used so-called fixed-node 
diffusion Monte Carlo for studying six electron quantum rings. 
They calculated also the lowest excited states for $M=0$. 
They found the ground state to have $S=0$ and the first excited state
$S=2$, in agreement with the CI calculations and the Hubbard model,
while for the second and third excited states they obtained the 
$S=1$ and $S=0$ states in opposite order as compared to the CI calculations.
Nevertheless, as the authors note, the energy differences in the narrow 
rings are so small that their difference starts to be within the 
statistical accuracy of the Monte Carlo method. 

The Monte Carlo studies for small quantum dots and rings show 
that while these methods can predict accurately the ground state energy,
they are not yet capable to give reliably the salient features of
the many-body spectrum.

\subsection*{Local density approximation}

The density functional Kohn-Sham method is another approach mainy suitable 
for the determination of the ground state structure. In applications
to quantum dots and rings (for a review see \cite{reimann2002}), the local
spin-density approximation (LSDA) is usually made and the system is assumed
to be strictly two-dimensional. Generally, the Kohn-Sham method is 
a 'mean field' method, where the electron-electron correlation is 
hidden in an effective single particle potential. This causes the
interesting feature that the mean field can exhibit symmetry 
breaking and the total electron and spin densities can reveal the 
{\it internal} symmetry of the ground state, for example the internal
shape of nucleus or atomic cluster\cite{hakkinen1997} or 
the static spin-density wave in a quantum dot\cite{koskinen1997}.
Indeed, in applications to quantum rings, the LSDA indicated 
the localization of electrons in an antiferromagnetic ring\cite{reimann1999}.
Nevertheless, we should add that although 
the LSDA can often eludicate the internal structure of a rotating system,
the method is not foolproof: In some cases, 
for example in rings or dots with high
enough electron density, it will not break the symmetry.

Systematic studies of quantum rings in terms of density functional
methods have been performed in several papers
\cite{emperador1999,emperador2000,viefers2000,emperador2001}, and
comparisons with 'exact' many-body methods show that the LSDA gives 
accurately the ground state energy\cite{pederiva2002}. 
The LSDA has been extended to so-called 
current-spin-density functional theory (CSDFT) 
which can take into account the gauge field\cite{vignale1988}.
Viefers {\it et al.} have applied the CSDFT for 
studying the persistent current, i.e. the ground
state energy as a function of the magnetic field, in small quasi-1D 
quantum rings. For a four electron ring (with $r_s=2.5$, $C_F=10$)
they found the yrast line consisting of two states: At zero flux
(and at $\Phi=\Phi_0$) the ground state had $S=1$ while around
$\Phi=\Phi_0/2$ the ground state had $S=0$. These results 
are in agreement with the CI calculations and the results of the 
Hubbard model for four electrons. Similar agreement was found 
for a six electron ring. The spin-densities showed a clear
localization of electrons in an antiferromagnetic ring.

Density functional theory has the same problem as quantum Monte Carlo
in that the determination of excited states is not straightforward
(although time-dependent current-spin-density-functional theory
can provide some information on 
excitations\cite{emperador1999,emperador2000,emperador2001}). 
Consequently, it is not possible to construct the complete excitation 
spectrum as by using the brute force CI method.

\section{Relation to Luttinger liquid}
\label{s130}

Infinitely long one-dimensional systems are 
often studied as Luttinger liquids\cite{luttinger1960,haldane1981}
(for reviews see \cite{haldane1994,voit1994,schulz1995}).
The speciality of the strictly 1D systems arises from the
fact that the Fermi surface consists of only two points
($\pm k_F$). This leads to a Peierls instability\cite{peierls1955}
and a breakdown of the Fermi liquid theory in a strictly
1D system. Important low energy excitations will then be 
collective, of bosonic nature, and have a linear dispersion relation.
The Luttinger liquid also exhibits so-called charge-spin 
separation: The spin and charge excitations can move with 
different velocities. In addition to studying the low energy
excitations, the Luttinger model has been extensively used 
for studying correlation functions\cite{voit1994,kolomeisky1996}.

It has been shown that the $tJ$-model is a Luttinger liquid\cite{voit1994}.
In the limit of a narrow ring with many electrons the spectral
properties of the quantum rings must then approach those of
a Luttinger liquid. We will now demonstrate that the many-body
spectra of quantum rings are consistent with the properties of
the Luttinger liquid. We do this only by qualitative considerations.
In an infinitely long 1D system, the low energy single particle excitations 
(of free fermions) are restricted to have a momentum change of $q\approx 0$
due to the fact that the Fermi surface is a point. 
In the Luttinger model it is precisely these excitations that
lead to the bosonization and collective plasmon excitations\cite{voit1994}
with a linear dispersion relation. In the case of a finite ring
these single particle excitations are just those described
in Fig. \ref{f1}, where one of the last electrons is excited from the 
compact state (or similarly in the Bethe ansatz solution of the $t$-model).
Now, it is exactly these single particle excitations which lead to the 
vibrational model, i.e. longitudinal acoustic phonons in the limit
of a long ring (see Fig. \ref{f4}), which have a linear dispersion 
relation (for small $q$). The excitation spectrum is in qualitative
agreement with the prediction of the Luttinger model already in the
smallest rings. 

Casting the Hamiltonian explicitly into charge dependent parts
(rotations and vibrations) and a spin-dependent part
(Heisenberg Hamiltonian),
as in Eqs.(\ref{e30_3}) and (\ref{eq:ModHam}) is equivalent to the charge-spin
separation in the Luttinger model. In an infinite system
this can be explicitly done for the half-full Hubbard model
($L=N$). We have demonstrated in Sec. \ref{s80} that the spin degrees
of freedom can be described with a good accuracy with 
the Heisenberg model in a much larger variety of quasi-1D rings.

\section{Pair correlation}
\label{s140}

The internal structure of a many-body electron state, especially
the possible localization to a Wigner molecule, can be
studied by examining the correlation functions. We have 
done this already for non-interacting electrons in Fig.
\ref{f3} where we used the $N$-particle correlation
function for identifying the vibrational states.
The $N$-particle correlation function is just the
square of the normalized many-body wave function.
A related analysis of the maximum $N$-particle correlation
for the Hubbard model, in Table \ref{t1}, also revealed
the internal stucture of the vibrational states.

The pair correlation function is frequently used to 
study the internal structure of a many-body state.
In one-dimensional systems the pair-correlations have the 
property that they decay with distance 
as $1/r^\alpha$\cite{kolomeisky1996}.
Consequently, the 1D electron system does not 
have a Wigner crystal with true long-range order.
The same is true for the spin-density oscillations: For 
example in the antiferromagnetic Heisenberg model
in 1D the spin-spin correlation decays as $1/r$,
$r$ being the distance between the electrons.

\begin{figure}
\includegraphics[angle=-90]{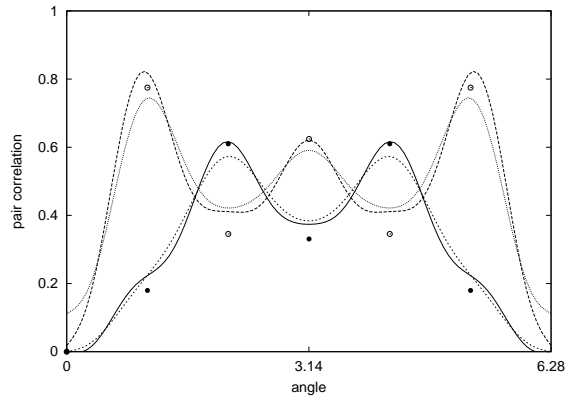}
\caption{Pair correlation function for six electrons in a quasi-one-dimensional
continuum ring, calculated with the CI method. Solid and long-dashed lines:
$\uparrow\downarrow$-correlation and $\uparrow\uparrow$-correlation, 
respectively, 
for a narrow ring with $r_s=2$ and $C_F=25$. Dashed and dotted lines:
$\uparrow\downarrow$-correlation and $\uparrow\uparrow$-correlation, 
respectively, 
for a wider ring with $r_s=2$ and $C_F=4$. The filled and open circles 
show the 
correlations for a six electron antiferromagnetic Heisenberg ring.
The functions are normalized so that the integral of the
$\uparrow\downarrow$-correlation
is 3 and that of the $\uparrow\uparrow$-correlation is 2.
}
\label{f19}
\end{figure}

In finite 1D rings, with only a few electrons,
the pair correlation function is even less informative.
The reason is again the Pauli exclusion principle, which
prevents electrons with the same spin to be at the same site. 
This means that within a short distance, the pair correlation 
functions are quite insensitive on the electron-electron
interaction. This is demonstrated in Fig. \ref{f19} where
we show the calculated pair correlation function for 
two different continuum rings\cite{koskinen2003} and compare
them to the pair correlation of the Heisenberg model.
We have also studied the pair correlation for four
electrons using the $t$-model and found that the correlation is
independent of the number of empty sites, as expected from the 
notion (Section \ref{s80}) that the spin-spin correlation is 
determined from the Heisenberg model, whatever the number of 
empty sites. 

The effect of temperature on the pair correlation function
has been studied by Borrmann and Harting\cite{borrmann2001} using 
quantum Monte Carlo. 
Koskinen {\it et al.}\cite{koskinenp2002} used the 
model Hamiltonian (\ref{eq:ModHam}) to determine the temperature 
dependence by calculating separately the pair correlation
for each quantum state. Both methods agree in the fact that 
the correlations between the electrons vanish as soon as the 
temperature exceeds the first excited state of the system.

\section{Interaction of the spin with the magnetic field: The Zeeman effect}
\label{s150}

Throughout most of this paper we have assumed that the magnetic flux is 
confined inside the ring so that the electrons move in a field-free
region. Experimentally, however, it might be difficult to produce a situation
where the magnetic field is zero at the ring site (or the
effective Land\'e factor is zero). It is then important to
consider also the Zeeman effect when comparing  
theory with experiments.
The interaction between the electron spin and the magnetic 
field adds to the Hamiltonian a term $\mu_Bg_0S_zB$.
In the case of a narrow ring it is beneficial
to write this as\cite{koskinenp2002}
\begin{equation}
H_Z=\frac{\hbar^2}{2m_eR^2}\left(\frac{\Phi}{\Phi_0}\right)gS_z,
\label{e1401}
\end{equation}
where we have written the field with help of the flux, 
ring radius, and an effective Land\'e factor $g$. 
For example, in the case of a homogeneous magnetic field and
an ideal 1D ring $\Phi=\pi R^2B$, and $g=g_0$. 
The advantage of writing the Zeeman part of the Hamiltonian 
as above is that we can study continuously the change in the 
spectra when we move from the case (\ref{noa1}), no Zeeman effect,
to a homogeneous magnetic field simply by changing $g$.
Moreover, the fact that the effective Land\'e factor in semicondictors
differs from that of the free electron can also be taken 
into account by just changing $g$. 

The periodicity of the persistent current, or lowest energy state,
as a function of flux is caused by the ground state energy jumping from
one angular momentum state to another. If all possible angular momenta
are visited ($J_{\rm eff}$ is small in the models discussed in Sec. \ref{s80}),
the periodicity is $\Phi_0/N$; if angular momenta $M=0,\, N/2, \, N$
etc. 
(or $N/2, \, 3N/2$ etc.)
are visited the periodicity is $\Phi_0/2$, and if
only angular momenta $M=0,\, N, \, 2N$ etc. are visited the 
periodicity is $\Phi_0$. In order to determine the overall periodicity
it is thus sufficient to examine the angular momentum values of the 
lowest energy state as a function of the flux.

\begin{figure}
\includegraphics{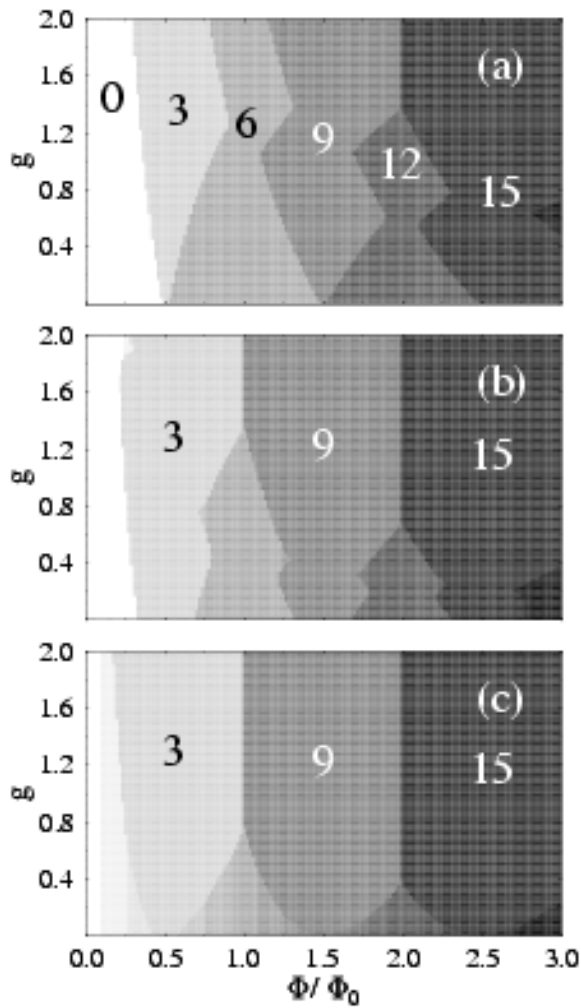}
\caption{Angular momentum of the ground state of a six
electron ring as a function of the flux and the effective
Lande factor $g$. The angular momentum is shown as numbers and
it increases with the darkness of the grayscale. The uppermost 
panel is for $JI=6$ in the model Hamiltonian (\ref{eq:ModHam}), 
the center panel for $JI=1.8$ and the lowest panel
for $JI=0$.}
\label{f23}
\end{figure}

Figure \ref{f23} shows the angular momentum of the ground state as 
a function of the flux and the effective Land\'e factor discussed
above. The results are shown for three different values of the 
parameter $J_{\rm eff}$. 
When the Land\'e factor approaches the free electron value $g=2$
and the flux is large, the periodicity always becomes $\Phi_0$.
The reason is the 
large Zeeman effect which makes the maximum spin state $S_z=N/2$
the lowest energy state. This state has always 
the periodicity $\Phi_0$ as shown in Fig. \ref{f9}.
In a wide ring (a) where the periodicity is $\Phi_0$ for $g=0$
it changes first to $\Phi_0/2$ and then again to $\Phi_0$
when $g$ increases. This is due to the fact that (in a ring
of six electrons) angular momenta $M=0,\, 6, \cdots$  are the 
ground states for the nonpolarized case while for the 
polarized case the ground state angular momenta are 
$M=3,\, 9,\cdots$. In the transition region a periodicity
$\Phi_0/2$ is observed. In the narrow ring (c) the periodicity 
changes smoothly from $\Phi_0/N$ to $\Phi_0$ when $g$ is 
increased from zero.

\section{Effect of an impurity}
\label{s160}

There has been extensive research on the
effects of impurities on the persistent current.
Already in 1988, Cheung {\it et al.} \cite{cheung1988} used a simple 
tight binding model for studying the effect of disorder in 1D rings
and found the decrease of the persistent current 
with increasing disorder.
Chakraborty and  Pietil\"ainen\cite{chakraborty1995}
used the CI method for studying the effect of an impurity 
in a polarized electron ring. The basic results of the 
included Gaussian impurity potential were to lift some
of the degeneracies of the energy levels and decrease 
the persistent current. Similar findings were observed by
Halonen {\it et al.}\cite{halonen1996} in the case where 
the the spin was included.
The suppression of the persistent current due to a gaussian
impurity was also computed by Viefers {\it et al.}\cite{viefers2000}
using density functional theory, thus including the effects of both 
spin, realistic interactions and finite width of the ring.
Eckle {\it et al.}\cite{eckle2000,eckle2001} have
used the Hubbard model to
study the effect of coupling the quantum ring 
to a quantum dot at Kondo resonance.
They have shown that at certain values of the flux the 
problem can be solved exactly using the Bethe ansatz.

\begin{figure}
\includegraphics[angle=-90]{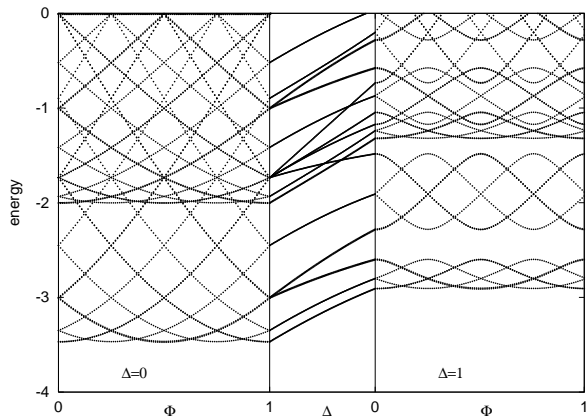}
\caption{Effect of an impurity on the energy levels of a
six-site Hubbard ring with four electrons. The left panel shows the 
energy levels as a function of the flux (in units of $\Phi_0$) for
an impurity free ring. The middle panel shows the evolution of the 
energy levels at zero flux as a function of an impurity potential
$\Delta$ at one of the lattice sites. The right panel shows the energy 
levels as a function of flux in the case where $\Delta=1$. 
$U=1000$ in all cases.
}
\label{f22}
\end{figure}

The effect of an impurity is easy to study in the
Hubbard model, where in the simplest case 
we can just add a repulsive or attractive 
potential in one of the lattice sites.
Figure \ref{f22} shows how the flux dependence of the 
energy levels changes when a repulsive external potential is 
introduced in one of the lattice sites. 
The effect is to open gaps in the spectrum.
The impurity potential splits the degeneracy of 
angular momenta (modulo 4) and makes all the energy levels
oscillate with a period of $\Phi_0$ instead of the 
period of $NL\Phi_0$ seen in the impurity-free case.
Note, however, that the levels corresponding to different 
angular momenta still cross and the lowest energy state has the same
period as without the impurity.

\begin{figure}
\includegraphics[angle=-90]{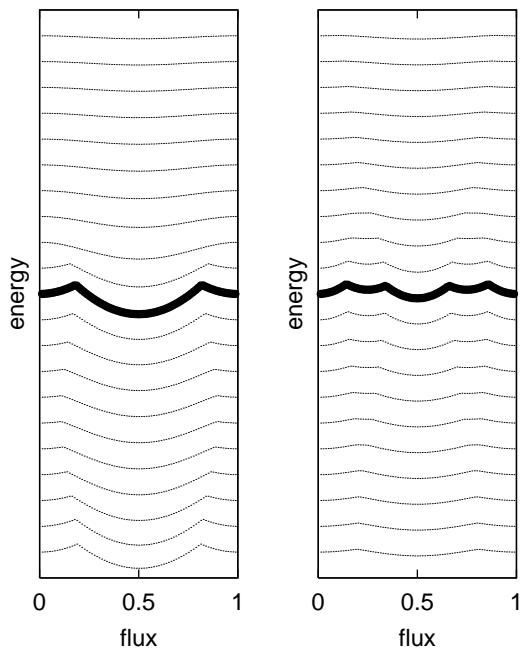}
\caption{Effect of an impurity on the lowest energy state in a 
Hubbard ring with $L=6$ and $N=4$. The left panel is for $U=20$ and the 
right panel for $U=200$. The thick line is the result without the impurity
and the other lines correspond to rings where the energy of one lattice site 
is increased (see text) by 2, 4, 6, etc. (dotted lines above the thick line) or
decreased by 2, 4, 6 etc. (dotted lines below the thick line). All lines 
have been shifted vertically so that they appear to be equally spaced. 
The energy scale in each line is the same, but it is twice as large for
$U=200$ as compared to $U=20$. The flux is in units of $\Phi_0=h/e$.
}
\label{f21}
\end{figure}

In the above example, the energy of the lattice
site, say site number 1, was changed by introducing a term
$\Delta (\hat n_{1\uparrow}+\hat n_{1\downarrow})$ in the 
normal Hubbard Hamiltonian, Eq. (\ref{eq:hub}). 
Figure \ref{f21} shows how the lowest energy state
as function of flux changes when
$\Delta$ is
varied from -20 to 20 in steps of 2.
For a relatively small value of $U$ (=20) and an attractive impurity,
the variation of the ground state
energy as a function of the flux is nearly independent of the 
impurity potential, while a positive (repulsive) impurity potential 
reduces the
amplitude of the oscillation, as expected from all earlier
studies\cite{chakraborty1995,niemela1996,halonen1996,viefers2000}.
Since the ground state persistent current is essentially the derivative
of the energy wrt. flux, see Eq.(\ref{ni10}), this directly reflects
the suppression of the current.
In the case of a relatively large $U$ (in the present example $U=200$),
we see that the impurity potential 
changes the original periodicity $\Phi_0/N$ to $\Phi_0/2$.

\begin{figure}
\includegraphics[angle=-90]{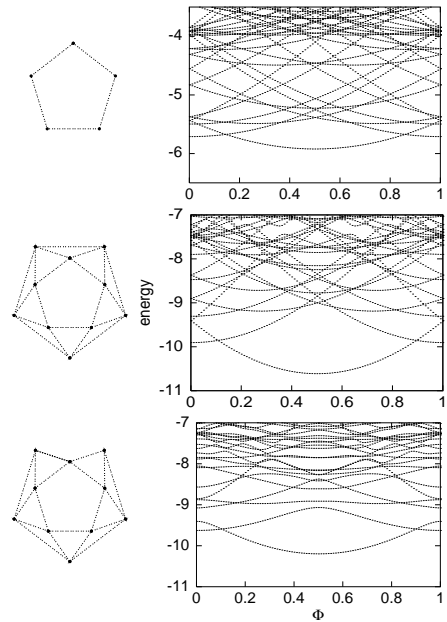}
\caption{Energy spectra as a function of flux for four
electron clusters. Upper panel: Hubbard ring with five sites and finite
$U=7$. Middle panel: ``Quasi-one-dimensional'' $t$-model ring.
Lowest panel: Quasi-one-dimensional $t$-model ring with a 
defect. The geometries in each case are shown at the left.
}
\label{f18}
\end{figure}

In the case of small systems the Hubbard model can easily be 
to used in studying also other kinds of impurities.
For example, two impurities or even random potentials at
lattice sites have qualitatively the same effect on a
Hubbard ring as a single impurity, as shown in Fig. \ref{f22}.
The Hubbard model may furthermore be used to study the coupling of the ring
to a quantum dot\cite{eckle2001}, as mentioned earlier.
In Figure \ref{f18} we present another example demonstrating how
the Hubbard model can be extended to quasi-1D rings
with impurities. In the uppermost panel of the figure we show the
energy spectrum of a 1D Hubbard ring with four electrons and a
finite $U=7$. The middle panel represents a {\it quasi}-1D ring 
with infinite $U$ ($t$-model). Comparison of these two
models shows that a finite $U$ in a strictly 1D Hubbard model 
actually mimics well the quasi-one-dimensionality of a more realistic ring,
as noticed already in comparing the Hubbard model with the 
continuum models.
The lowest panel of Fig. \ref{f18} shows an 'impurity' or
a narrow neck in the ring. Its effect is to open gaps
in the excitation spectrum. Qualitatively, the effect is the same 
as observed above for a 1D Hubbard ring with an impurity potential.

In many experiments, the persistent current is measured for a
collection of rings, possibly
consisting of several propagating channels (radial modes). 
The persistent currents in such rings are not
determined by quantum mechanical eigenenergies of a single ring, 
but are affected by several complications like disorder and ensemble
averaging \cite{vonoppen1991,altshuler1991,montambaux1990}.
The electron-electron interactions seem to play a crucial role
also in the disordered rings\cite{ambegaokar1990}.
M\"uller-Groeling and Weidenm\"uller have shown that interactions
effectively counteract the impurity suppression of the persistent 
current\cite{muller1994},
in qualitative agreement with our Hubbard model results.

\section{Summary}
\label{sum}

In this paper we have attempted to present a comprehensive
review of the physics of few-electron quantum rings, with
particular focus on their energy spectra and the periodicity
of the persistent current. We compared various analytical and numerical
theoretical approaches which fall into two main classes -- 
lattice models on the
one hand and continuum models on the other -- and tried to 
clarify the connections between them. The main message is that
all the different approaches give essentially the same results
for the spectra and persistent currents
(provided one takes the continuum limit of the discrete models
to compare them to the continuum ones).
The essential physics is captured by a simple model Hamiltonian
which describes the many-body energy as a combination of rigid
rotation and internal vibrations of the ring `molecule', plus a
Heisenberg term which determines the spin dynamics. In the spectra, 
the vibrational excitations can be seen as higher bands, while
the lowest (yrast) band is purely rotational. Its periodicity with
respect to the flux is always $\Phi_0$ in the case of spinless 
(polarized) electrons, while for a clean ring of nonpolarized 
electrons, the periodicity changes from $\Phi_0$ via $\Phi_0/2$
to $\Phi_0/N$ as the ring get narrower or, in the language of
the Hubbard model, as the interaction strength $U$ increases
(the effectvie Heisenberg coupling $J$ decreases).
Impurities in the ring may change the periodicity (to $\Phi_0$)
even in the nonpolarized case, and moreover suppress the
persistent current.

\subsection*{Acknowledgements}
We would like to thank M. Koskinen for enlightening discussions.
This work has been supported by Nordita and by the Academy of Finland under
the Finnish Centre of Excellence Programme 2000-2005 (Project No. 44875,
Nuclear and Condensed Matter Programme at JYFL).


\begin{thebibliography}{99}
\bibitem{footnote1} In fact, the Aharonov-Bohm phase 
can be regarded as a special case
of the Berry phase.
\end{thebibliography}
\end{document}